\documentclass{ar-1col-S2O}

\usepackage[numbers]{natbib}
\usepackage{url}
\usepackage{amsmath}
\usepackage{amssymb}
\usepackage[
            linkcolor=blue,
            citecolor=blue,
            urlcolor=blue]{hyperref}

\jname{Submitted to Annual Review of Condensed Matter Physics}
\jyear{2026}

\begin{document}

\markboth{Wang}{Strong-to-Weak SSB}

\title{Strong-to-Weak Spontaneous Symmetry Breaking}

\author{Chong Wang$^1$
\affil{$^1$Perimeter Institute for Theoretical Physics, Waterloo, Ontario N2L 2Y5, Canada; email: cwang4@pitp.ca}
}

\begin{abstract}
Strong-to-weak spontaneous symmetry breaking (SW-SSB) has recently emerged as a useful framework for studying phases of matter in open systems, quantum or classical. Beginning with the simple idea of extending symmetry breaking to general mixed states, and the familiar equivalence between canonical and grand-canonical ensembles in statistical mechanics, the concept has grown into a unifying perspective connecting many different ideas in physics, including topological orders, emergent hydrodynamics, and information-theoretic characterization of phases of matter. This review provides a bird’s-eye view of some of these recent developments.
\end{abstract}

\maketitle

\tableofcontents

\section{INTRODUCTION}

It is perhaps not an exaggeration to say that the history of modern theoretical physics is largely an endless quest to better understand symmetries. In both condensed matter and high-energy physics, the notion of spontaneous symmetry breaking (SSB) provides the foundation for classifying phases of matter, from laboratory materials (such as solids, magnets and superconductors) to the very universe we live in.~\cite{LandauLifshitz,McGreevy2023}

Recently, the theory of a new form of symmetry breaking, dubbed strong-to-weak SSB (SW-SSB), has been developed~\cite{GarrattChalker2021,LeeJianXu2023,OgunnaikeFeldmeierLee2023,Moudgalya_2024,ChenGrover,Maetal2025,Lessa2024,sala_spontaneous_2024}. This phenomenon arises naturally in open quantum systems, where the states are described by density matrices $\rho$. The same notion also applies to classical stochastic systems described by probability distributions. This perspective opens up new angles for exploring phases of matter, which is the focus of this review.

Let us begin by recalling how symmetry acts in a quantum system. A pure quantum state $|\psi\rangle$ is symmetric under a unitary symmetry $g$ if $g$ is represented by a unitary operator $U_g$ such that
\begin{equation}
U_g|\psi\rangle=e^{i\alpha}|\psi\rangle,
\end{equation}
where the eigenvalue $e^{i\alpha}$ is the symmetry charge of $|\psi\rangle$. For mixed states, there are two ways to generalize the symmetry condition. The most common symmetry condition is
\begin{equation}
\label{eq:weaksymm}
U_g\rho U_g^{\dagger}=\rho.
\end{equation}
A symmetry $g$ satisfying this condition is called a \textbf{weak symmetry} of $\rho$. For example, a Gibbs state $\rho=e^{-\beta H}/Z$ has $g$ as a weak symmetry whenever $g$ is also a symmetry of the Hamiltonian, i.e., $[U_g,H]=0$.

However, one may impose a stronger symmetry condition
\begin{equation}
\label{eq:strongsymm}
U_g\rho=e^{i\alpha}\rho,
\end{equation}
which defines what is known as a \textbf{strong symmetry}~\cite{BucaProsen2012,AlbertJiang2014}. To understand its meaning, it is useful to view the mixed state as an ensemble of pure states,
\begin{equation}
\label{eq:decomposition}
\rho=\sum_I P_I|\psi_I\rangle\langle\psi_I|,
\end{equation}
where $\sum_I P_I=1$ and $P_I\ge 0$ represents the probability of finding the system in the pure state $|\psi_I\rangle$. Eq.~\ref{eq:decomposition} is called a ``convex decomposition.'' The states $|\psi_I\rangle$ in a convex decomposition need not be orthogonal to each other, so the decomposition is generally not unique for a genuine mixed state.

A simple yet instructive exercise in linear algebra shows that the strong symmetry condition Eq.~\ref{eq:strongsymm} implies that $U_g|\psi_I\rangle=e^{i\alpha}|\psi_I\rangle$ for any $|\psi_I\rangle$ in any convex decomposition. In other words, every pure state in the ensemble must itself be $g$-symmetric with the same symmetry charge.

By contrast, the weak symmetry condition Eq.~\ref{eq:weaksymm} does not guarantee any individual pure state to be symmetric -- the symmetry is only a property of the statistical ensemble, and is therefore sometimes called an ``average symmetry''. While weak symmetry does imply the existence of a convex decomposition in terms of symmetric pure states $|\psi_I^{\text{sym}}\rangle$, the symmetry charges can differ among them: $U_g|\psi_I^{\text{sym}}\rangle=e^{i\alpha_I}|\psi_I^{\text{sym}}\rangle$ ($\alpha_I$ depends on $I$).

With these two notions of symmetry in mind, let us revisit familiar examples of spontaneous symmetry breaking. Consider first a classical Ising ferromagnet,
\begin{equation}
\label{eq:classicalmagnet}
    \rho=\frac{1}{2}|\uparrow\uparrow\uparrow...\rangle\langle\uparrow\uparrow\uparrow...|+\frac{1}{2}|\downarrow\downarrow\downarrow...\rangle\langle \downarrow\downarrow\downarrow...|,
\end{equation}
which is a classical mixture (a flipped coin). By contrast, a quantum Ising ferromagnet can be described by a pure state
\begin{equation}
    \label{eq:quantummagnet}
    |\Psi\rangle=\frac{1}{\sqrt{2}}|\uparrow\uparrow\uparrow...\rangle+\frac{1}{\sqrt{2}}|\downarrow\downarrow\downarrow...\rangle,
\end{equation}
known as the Greenberger–Horne–Zeilinger (GHZ) state, or ``cat state.'' In both cases, a realistic observer will find either all spins up or all spins down, indicating complete breaking of the $\mathbb{Z}_2$ spin-flip symmetry. When the state collapses to a definite symmetry-breaking configuration, such as $|\uparrow\uparrow\cdots\rangle$ or $|\downarrow\downarrow\cdots\rangle$ after a measurement, we say that the symmetry (whether strong or weak) has been broken explicitly at the level of selected states.

However, the symmetry is realized differently in the two cases. The $\mathbb{Z}_2$ symmetry, represented by $\otimes_i X_i$ ($X_i$ being the Pauli-$X$ operator at site $i$), is a strong symmetry for the quantum magnet, since the state is pure. In contrast, for the classical magnet it is only a weak symmetry, as the individual pure states in the ensemble are not themselves symmetric. Thus, the classical magnet exhibits a symmetry-breaking pattern of weak $\to$ none, while the quantum magnet exhibits strong $\to$ none. This naturally raises the question: can a symmetry-breaking pattern from strong to weak occur? If so, what is the physical meaning? What are its consequences?

The notion of strong-to-weak symmetry breaking can also be motivated from a different angle. The distinction between strong and weak symmetry is reminiscent of the familiar canonical and grand-canonical ensembles in statistical physics. In a system with particle-number conservation (a $U(1)$ symmetry), a canonical ensemble consists of states with fixed total charge (corresponding to a strong $U(1)$ symmetry), whereas a grand-canonical ensemble consists of states with different total charges weighted by a chemical potential (corresponding to a weak $U(1)$ symmetry). The physical origin of the two ensembles lies in whether the system exchanges charge with the environment. The same intuition applies more generally: a weak symmetry typically arises from charge exchange between the system and its environment, while a strong symmetry requires that the system–environment coupling does not exchange symmetry charge.

A fundamental statement in equilibrium statistical physics is the equivalence between the canonical and grand-canonical ensembles: although not identical, they are operationally indistinguishable in the thermodynamic limit. In other words, even if the strong symmetry condition is imposed (canonical ensemble), the system behaves as if only the weak symmetry were present (grand-canonical ensemble). Viewed as an effective reduction of symmetry from strong to weak, this equivalence already resembles a form of spontaneous symmetry breaking. In recent years, this idea has been formalized as ``strong-to-weak spontaneous symmetry breaking'' (SW-SSB). Since the equivalence between statistical ensembles is a hallmark of thermal equilibrium, it is natural to expect a deep connection between SW-SSB and thermalization. Indeed, early work has linked thermalization to the spontaneous breaking of strong time-translation symmetry~\cite{GarrattChalker2021}. More recently, this perspective has been extended to general symmetries, leading to the understanding that SW-SSB provides a sharp measure of how much the symmetry charge is spread globally~\cite{Lessa2024}. Even when the state is not thermal, i.e. it does not locally resemble a Gibbs state, the occurrence of SW-SSB still indicates a form of ``charge thermalization''.

There is yet another line of development relevant to us: characterizing many-body states from an information-theoretic point of view. Traditional spontaneous symmetry breaking is typically diagnosed through connected two-point correlation functions. From an information-theoretic perspective, such correlations can be captured through \textit{mutual information} (MI), which measures how much information about one region can be inferred from another distant region. More recently, it has been appreciated that a subtler form of correlation, described by \textit{conditional mutual information} (CMI), is equally important~\cite{sang2024stability}. Roughly speaking, CMI quantifies correlations between two regions that are not mediated by the rest of the system. In this language, conventional SSB provides the simplest example of long-range mutual information, whereas SW-SSB provides the simplest example of long-range conditional mutual information~\cite{Lessa2024}.

The development of SW-SSB is part of a broader effort to understand many-body physics in open quantum systems~\cite{hastings2011nonzero,coser2019classification,de2022symmetry,ma2023average,FanBaoMixedTopo,LeeYouXu2022,ChenGrover2,sang2023mixed,StableOpen2024,sang2024stability,lessa2025higher,Sang2025LR} that are not necessarily thermal. This topic is timely: modern quantum simulators inevitably couple to their environment, yet remain sufficiently controllable to preserve coherence over meaningful timescales. Understanding the landscape of many-body quantum phases in the era of noisy intermediate-scale quantum (NISQ) devices is therefore of clear current interest. Recently, the first experimental observation of SW-SSB was reported in a dephased cold Fermi gas~\cite{Experiment2026}, marking an important step in the development of the field and transforming SW-SSB from a purely theoretical concept into an experimentally accessible phenomenon.

The theory of SW-SSB exemplifies the unity of physics. Ideas from traditionally different viewpoints -- spontaneous symmetry breaking, thermalization, and (quantum) information theory -- blend together naturally. The result is a useful framework for characterizing mixed-state phases -- classical or quantum, thermal or non-thermal -- and has led to the identification of new phases and phase transitions. The concept also connects naturally to several active directions in modern physics, ranging from topological phases in open quantum systems to the emergence of hydrodynamics in charge-conserving systems. This review aims to highlight this harmonious interplay among different ideas and to provide a pedagogical account of these developments. Given that the subject is still under active exploration, we focus primarily on the most basic conceptual aspects, and conclude with a brief suggestive outlook.

\section{THE ORDER PARAMETERS}

\subsection{Fidelity correlator and the stability theorem}

Ordinary symmetry breaking is typically measured by some local order parameter. For example, consider a pure state $|\psi\rangle$, then symmetry breaking can be measured by
\begin{equation}
\label{eq:purestateorder}
    \langle O_i\rangle=\langle\psi|O_i|\psi\rangle,
\end{equation}
where $O_i$ is a local operator acting near position $i$, and is charged under some symmetry operator $g$, such that $U_g O_i U^{\dagger}_g=e^{i\theta}O_i$ where $e^{i\theta}$ is the charge carried by $O_i$ -- for Ising magnets we can simply consider $O_i=Z_i$ (the Pauli-$Z$ operator). If the state does not break the symmetry explicitly -- for example the cat state in Eq.~\ref{eq:quantummagnet} -- then $\langle O_i\rangle=0$ and we can detect the spontaneous breaking of the symmetry through the two-point correlator
\begin{equation}
    \label{eq:puretwopt}
    \langle O_{ij}\rangle :=\langle\psi|O_i O_j^{\dagger}|\psi\rangle,
\end{equation}
where $i$ and $j$ are two far-separated points in space. 

The order parameter in Eq.~\ref{eq:purestateorder} (or its two-point version Eq.~\ref{eq:puretwopt}) can be interpreted in two ways (below $O$ can be either $O_i$ or $O_{ij}=O_iO_j^{\dagger}$):
\begin{enumerate}
\item it is the expectation value obtained by measuring the operator $O$;
\item it quantifies the overlap, or similarity, between the original state $|\psi\rangle$ and the state obtained by acting with the charged operator, $O|\psi\rangle$.
\end{enumerate}

For pure state, the above two interpretations are completely equivalent -- in fact this equivalence is one reason to view Bose-Einstein condensation as a form of spontaneous symmetry breaking~\cite{Anderson1966}. Things become slightly subtle when generalized to mixed states. The standard generalization is
\begin{equation}
    \langle O\rangle ={\rm tr}\rho O,
\end{equation}
which admits the first interpretation, but not the second. If we want to quantify the similarity between the original state $\rho$ and the state obtained by acting with $O$, namely $O\rho O^{\dagger}$, we should consider the overlap between the two matrices given by:
\begin{equation}
\label{eq:FC}
    \langle O\rangle_F:=F(\rho,O\rho O^{\dagger}),
\end{equation}
where $F$ is the Uhlmann fidelity commonly used to measure similarity between two density matrices:
\begin{equation}
    F(\rho,\sigma):={\rm tr}\sqrt{\sqrt{\rho}\sigma\sqrt{\rho}}.
\end{equation}
Eq.~\ref{eq:FC} is called the \textbf{fidelity correlator} in Ref.~\cite{Lessa2024} and it can be either one-point ($O=O_i$) or two-point ($O=O_iO_j^{\dagger}$). For a pure state, $\langle O\rangle_F=|\langle O\rangle|$. One can also show that for general mixed states $\langle O\rangle_F\geq |\langle O\rangle|$.

So what do all these have to do with symmetry breaking? Recall that the reason $\langle O\rangle\neq0$ signals usual symmetry breaking is that symmetry requires $\langle O\rangle=0$. More specifically, weak symmetry requires
\begin{equation}
    {\rm tr}\rho O_i={\rm tr}U_g\rho U_g^{\dagger} O_i=e^{-i\theta}{\rm tr}\rho O_i,
\end{equation}
which in turn requires the order parameter to vanish for any $e^{i\theta}\neq1$. Naturally, one expects that if the state $\rho$ is also strongly symmetric, a stronger result should follow. Indeed, since strong symmetry requires all the pure states $|\psi_I\rangle$ in $\rho=\sum_I P_I|\psi_I\rangle\langle\psi_I|$ to have the same symmetry charge $e^{i\alpha}$, any pure state in $\sigma=O_i\rho O_i^{\dagger}$ will have a different charge $e^{i(\alpha+\theta)}$, and is therefore orthogonal to any pure state in $\rho$. Hence, strong symmetry implies
\begin{equation}
\langle O_i\rangle_F=F(\rho,O_i\rho O_i^{\dagger})=0.
\end{equation}
Since $\langle O\rangle_F\geq|\langle O\rangle|$, this statement is stronger than that implied by weak symmetry, as expected.

We can then use the one-point fidelity correlator $\langle O_i\rangle_F$ to detect the explicit breaking of a strong symmetry. If the state preserves the strong symmetry, $\langle O_i\rangle_F=0$. We can similarly define spontaneous breaking of a strong symmetry as the non-vanishing of the two-point fidelity correlator $\langle O_i O_j^{\dagger}\rangle_F$ as $|i-j|\to\infty$. If the strong symmetry is spontaneously broken in this way, but weak symmetry is not spontaneously broken, so that the ordinary correlation function vanishes $\langle O_i O^{\dagger}_j\rangle\to0$ as $|i-j|\to\infty$, then we say that strong-to-weak spontaneous symmetry breaking (SW-SSB) has occurred.

It is useful to have a simple example in mind. Consider a lattice system with one qubit on each site $i$, and an Ising $\mathbb{Z}_2$ symmetry represented by $X:=\otimes_i X_i$. The simplest state that exhibits $\mathbb{Z}_2$ SW-SSB is
\begin{equation}
    \rho_0=\frac{I+X}{2^N},
\end{equation}
where $N$ is the total number of qubits. This is nothing but a maximally mixed state within the $X=1$ eigenspace, or the ``infinite-temperature'' state in a ``canonical ensemble'' with fixed symmetry charge $X=1$. $\rho_0$ plays a prototypical role in SW-SSB, much like the classical flipped coin (Eq.~\ref{eq:classicalmagnet}) in weak-to-none SSB and the cat state (Eq.~\ref{eq:quantummagnet}) in strong-to-none SSB. Clearly the ordinary correlator $\langle Z_iZ_j\rangle=0$ for any $i\neq j$. The fidelity correlator is equally simple to calculate: since $Z_iZ_j\rho_0 Z_iZ_j=\rho_0$, we have $\langle Z_iZ_j\rangle_F=1$ for any $i,j$. So $\rho_0$ indeed exhibits SW-SSB as we defined.

The above example also demonstrates the physical meaning of SW-SSB: if the fidelity correlator $\langle Z_i Z_j\rangle_F$ is large (i.e., a non-decaying $O(1)$ constant), then moving the symmetry charge around (from $i$ to $j$) does not significantly change the state. This implies that the symmetry charge is highly delocalized throughout the system, and the total charge has become a piece of global information. In the case of $\rho_0$, the charge is completely randomly distributed in the entire system, and even tracing out one qubit destroys the information about the total charge, as it leads to a maximally mixed state on the remaining qubits:
\begin{equation}
    {\rm tr}_i\rho_0=\frac{I^{(N-1)}}{2^{N-1}}.
\end{equation}
SW-SSB may therefore be viewed as a form of ``charge thermalization'', with the quotation marks emphasizing that it may not be equivalent to any standard notion of thermalization.

\subsubsection{Symmetry-breaking perturbation}

Another hallmark of traditional SSB is its sensitivity to infinitesimal symmetry-breaking perturbation: starting from a long-range ordered state like Eq.~\ref{eq:classicalmagnet} or \ref{eq:quantummagnet} and turning on a symmetry-breaking field $h\to 0^+$, the system immediately collapse into an explicitly symmetry-breaking state $|\uparrow\uparrow\uparrow...\rangle$. SW-SSB has a similar feature under infinitesimal strong-symmetry-breaking quantum channel~\cite{LeeJianXu2023,zhang2024fluctuation}. 

Let us first review some basic facts about quantum channels and their symmetries. Consider a quantum channel, namely a completely positive and trace-preserving (CPTP) map, on density matrices $\mathcal{E}: \rho \to \mathcal{E}[\rho]$. Conceptually it is often appealing to work with the Stinespring dilation form:
\begin{equation}
    \mathcal{E}[\rho]={\rm tr}_eU\left(\rho\otimes|0_e\rangle\langle0_e|\right)U^{\dagger},
\end{equation}
where $|0_e\rangle$ is a state in an ancilla (environment) Hilbert space, $U$ is a unitary operator acting on both the system and the ancilla, and the ancilla is traced out at the end of the process. For a symmetry $g$ represented by a unitary $U_g$ acting on the system, the channel preserves the strong symmetry if $U$ commutes with $U_g\otimes I_e$, where $I_e$ is the identity operator acting on the ancilla -- physically, this simply means that the environment does not absorb charge from the system. 

One can also expand the dilation form in an ancilla basis to obtain the Kraus form of a quantum channel:
\begin{equation}
    \mathcal{E}[\rho]=\sum_aK_a\rho K_a^{\dagger},
\end{equation}
with the Kraus operators satisfying $\sum_a K_a^{\dagger}K_a=1$. In this form, the strong symmetry condition for the channel becomes $[K_a, U_g]=0$ for any $K_a$. 

In contrast, to preserve a weak symmetry, only a weaker condition needs to be satisfied by the channel:
\begin{equation}
    \mathcal{E}[U_g\rho U_g^{\dagger}]=U_g\mathcal{E}[\rho]U_g^{\dagger}, \hspace{20pt}\forall\rho.
\end{equation}

For concreteness, let us again consider an Ising system with strong $\mathbb{Z}_2$ symmetry $X:=\otimes_iX_i$. The simplest quantum channel that breaks the strong symmetry but keeps the weak one is
\begin{equation}
\label{eq:weakmeasure}
    \mathcal{E}_p=\otimes_i\mathcal{E}_i, \hspace{20pt} \mathcal{E}_i[\rho]=(1-p)\rho+pZ_i\rho Z_i,
\end{equation}
which can be interpreted either as a weak measurement of $Z$ on all sites $i$, or a unitary $Z_i$ acting on the system with probability $p$. The parameter $p>0$ also measures the degree of strong symmetry breaking by the channel. 

If we start from a strongly symmetric state $\rho$, then by strong symmetry $\langle  Z_i\rangle^{\rho}_F=0$. Now acting with the channel Eq.~\ref{eq:weakmeasure}, the strong symmetry is broken and a nonzero $\langle Z_i\rangle^{\mathcal{E}[\rho]}_F$ can be induced. So what does SW-SSB imply here? To gain some intuition, consider the simplest initial state with SW-SSB, namely $\rho_0\sim I+X$. A simple calculation shows that for any $p>0$ (even if very small), in the large volume limit $\mathcal{E}_p[\rho_0]\sim I$ which is just the maximally mixed state, with $\langle Z_i\rangle_F=1$. Such sensitivity to infinitesimal symmetry-breaking perturbations is another way to characterize spontaneous symmetry breaking. To be more precise, we can define a (strong-to-weak) version of susceptibility, dubbed ``fidelity susceptibility'' in Ref.~\cite{zhang2024fluctuation}:
\begin{equation}
    \chi_F^{\rho}=\lim_{p\to 0^+}\frac{\langle Z_i\rangle_F^{\mathcal{E}_p[\rho]}-\langle Z_i\rangle_F^{\rho}}{p}.
\end{equation}
It can be proved~\cite{zhang2024fluctuation} (with some mild assumptions such as weak translation symmetry) that this susceptibility is bounded from above and below by the fidelity correlator:
\begin{equation}
    \sum_j\left(\langle Z_iZ_j\rangle_F\right)^2\leq\frac{1}{4}\chi_F\leq\left(\sum_j\langle Z_iZ_j\rangle_F \right)^2.
\end{equation}
This is reminiscent of the standard \textit{fluctuation-dissipation theorem} connecting susceptibility to correlation functions. This relation guarantees that, for SW-SSB states with non-decaying two-point fidelity correlator, $\chi_F$ diverges with system size; while for non-SSB states with exponentially decaying fidelity correlator, $\chi_F$ remains finite in the thermodynamic limit.

\subsubsection{Hidden Edwards-Anderson order}

The fidelity correlator has a curious relation to the Edwards-Anderson (EA) order parameter familiar in spin glass physics~\cite{EdwardsAnderson}. For a statistical ensemble of states $\{|\Psi_I\rangle,P_I\}$ (with probability $P_I$ to find the system in state $|\Psi_I\rangle$), the EA order parameter for an operator $O$ is the average of the absolute value
\begin{equation}
\label{eq:EA}
    \langle O\rangle_{EA}=\sum_IP_I|\langle\Psi_I|O|\Psi_I\rangle|.
\end{equation}
Physically, the EA parameter measures the symmetry-breaking order within each sample in the ensemble, even when the ordinary order parameters take random signs and average to zero in the entire ensemble, in which case the exact symmetry (for each sample) is broken to an average symmetry (for the whole ensemble). This picture clearly shares some similarity with SW-SSB. 

We again consider the simple mixed state $\rho_0\sim I+X$. We can write it as the convex sum of all cat states with symmetry charge $X=1$:
\begin{equation}
    \rho_0=\frac{I+X}{2^N}=\sum_{\{s\}}\frac{1}{2^{N-1}}|\Psi_{\{s\}}\rangle\langle\Psi_{\{s\}}|, \hspace{10pt} |\Psi_{\{s\}}\rangle=\frac{1}{\sqrt{2}}(\otimes_i|Z_i=s_i\rangle+\otimes_i|Z_i=-s_i\rangle),
\end{equation}
where $\{s\}=\{s_1,s_2...\}$ is a string of $N$ bits taking value $\pm1$. In this particular decomposition, it can be easily verified that the fidelity correlator $\langle Z_iZ_j\rangle_F$ is exactly the EA correlator~\cite{Lessa2024}.

In general, the value of the EA correlator for a mixed state, when viewed as a statistical ensemble of pure states, will sensitively depend on the choice of convex decomposition $\rho=\sum_IP_I|\Psi_I\rangle\langle\Psi_I|$. Indeed for $\rho_0\sim I+X$, we can choose another decomposition in $X$-basis 
\begin{equation}
    \rho_0=\frac{I+X}{2^N}=\sum_{\{s\},\prod_is_i=1}\frac{1}{2^{N-1}}\otimes_i|X_i=s_i\rangle\langle X_i=s_i|,
\end{equation}
which clearly has no EA order, since each pure state in the ensemble is a symmetric product state. 

So is there any intrinsic meaning to the EA order, given its dependence on convex decomposition? It turns out that for a strongly symmetric state $\rho$, a convex decomposition with long-range EA order exists if and only if $\rho$ has SW-SSB~\cite{ChenGrover}. In fact the fidelity correlator is precisely the maximal value of the EA correlator over all such convex decomposition~\cite{Lessatoappear}. This establishes a precise connection between SW-SSB in mixed states and EA order from spin glass physics. 

\subsection{Canonical purification and R\'enyi-$1$ correlator}

We used Uhlmann fidelity to define the correlator characterizing SW-SSB. There are many other measures of similarity between two different density matrices, such as trace distance and relative entropy\cite{LeeJianXu2023,Lessa2024,Knap2025}. These different measures of overlap lead to different quantities, and it turns out that they are all equivalent as definitions of SW-SSB\cite{Lessa2024,Knap2025}. A particularly interesting one is the Holevo fidelity $F_H(\rho,\sigma)={\rm tr}[\sqrt{\rho}\sqrt{\sigma}]$, from which one can motivate the R\'enyi-$1$ correlator, also known as the Wightman correlator\cite{LiuRenyi1,WeinsteinRenyi1}:
\begin{equation}
\label{eq:R1}
    \langle O\rangle_{R1}={\rm tr}[\sqrt{\rho}O\sqrt{\rho}O^{\dagger}].
\end{equation}

The R\'enyi-$1$ correlator admits an appealing interpretation in terms of \textit{canonical purification}. In general, a mixed state $\rho$ in a Hilbert space $\mathcal{H}$ admits infinitely many purifications $|\Psi\rangle\in \mathcal{H}\otimes\mathcal{H}^a$ ($\mathcal{H}^a$ is the ancilla and by definition ${\rm tr}_a|\Psi\rangle\langle\Psi|=\rho$). In the study of mixed-state phases of matter, canonical purification (also known as thermofield double) turned out to be particularly interesting. Essentially, one can view the matrix $\sqrt{\rho}$ as a vector in a doubled Hilbert space $|\sqrt{\rho}\rangle\rangle\in\mathcal{H}^L\otimes\mathcal{H}^R$, with $L$ and $R$ labeling left and right indices in the original matrix. More concretely, we can work with any choice of basis with $\sqrt{\rho}=(\sqrt{\rho})_{ab}|a\rangle\langle b|$ and convert to the doubled space $|\sqrt{\rho}\rangle\rangle=(\sqrt{\rho})_{ab}|a\rangle|b\rangle^*$ where $^*$ denotes complex conjugation (one can check that the final state obtained this way does not depend on basis choice). Viewing $\mathcal{H}^L$ as the original system and $\mathcal{H}^R$ as the ancilla, one can easily check that $|\sqrt{\rho}\rangle\rangle$ is indeed a purification. 

In the doubled-space formulation, we assign the ancilla the same locality structure as the system. Namely, for each lattice site $i$, the spins from both system and ancilla are regarded as occupying the same spatial location $i$. 

The meaning of strong vs. weak symmetry becomes quite transparent in the doubled space: strong symmetry acts only on one copy of the Hilbert space, as either $U^L_g\otimes I^R$ or $I^L\otimes (U^*_g)^R$, while weak symmetry acts simultaneously on both as $U^L_g\otimes (U^*_g)^R$. Formally, the strong symmetry $G$ becomes doubled $G^L\times G^R$ while the weak symmetry is the diagonal subgroup $G$.

The R\'enyi-$1$ correlator Eq.~\ref{eq:R1} can be expressed in an appealing form in the doubled space:
\begin{equation}
    \langle O\rangle_{R1}=\langle\langle\sqrt{\rho}|O^L(O^R)^{\dagger}|\sqrt{\rho}\rangle\rangle,
\end{equation}
which is an ordinary correlation function in the doubled space. Noticing that $O_i^L (O_i^R)^{\dagger}$ is charged under the strong symmetry but not the weak, so this correlation function is precisely probing the breaking of the strong $G^L\times G^R$ to the weak diagonal $G$ symmetry. As a concrete example, one can check explicitly that the simplest SW-SSB state $\rho_0\sim I+X$ corresponds to a doubled state $|\rho_0\rangle\rangle$ that is nothing but a cat state with long-range order.

The equivalence between the fidelity and R\'enyi-$1$ correlators as characterization of SW-SSB is guaranteed by the following inequalities, proved in Refs~\cite{LiuRenyi1,WeinsteinRenyi1}:
\begin{equation}
\label{eq:R1bounds}
    \langle O\rangle_F^2\leq \langle O\rangle_{R1}\leq \langle O\rangle_F ||O||_{\infty},
\end{equation}
where $||O||_{\infty}$ is the operator norm of $O$ (the largest singular value). 

It is quite curious that canonical purification can faithfully encode mixed-state symmetry-breaking order in terms of ordinary pure-state symmetry breaking. A similar feature appears in the study of topological phases -- both symmetry-protected topological (SPT) phases \cite{de2022symmetry,ma2023average,zhang2022strange,LeeYouXu2022,Maetal2025,Ma2024SPTdoubled,XueLeeBao,Guo2025,Lu2024,Manjunath2025} and intrinsic topological orders\cite{BaoFandouble,ChenGrover2,WangWuWang,RamanjitAbhinav,EllisonCheng} -- in mixed states: under canonical purification, a mixed-state topological phase maps to a corresponding pure-state topological phase in a somewhat faithful manner. This recurring pattern suggests more than a coincidence. In a recent work~\cite{Li2026} it was shown that a mixed state with ``short-range correlation'', in the sense of exponentially decaying mutual information and conditional mutual information (to be reviewed in Sec.~\ref{sec:CMI}), has canonical purification that is also short-range correlated. This suggests a deep connection between quantum phases in mixed and pure states~\cite{Li2026,Huangtoappear}.

\subsection{Choi double and R\'enyi-$2$ correlator}

Both the Uhlmann and Holevo fidelity require taking $\sqrt{\rho}$, which is in general difficult to compute even if $\rho$ is given (see Ref.~\cite{LiuZou2026} for some recent progress on bounding such quantities in matrix product states). A more computable measure of overlap between two matrices is the Frobenius inner product ${\rm tr}(\rho\sigma)$. The Frobenius product motivates the R\'enyi-$2$ correlator for SW-SSB\cite{LeeJianXu2023}:
\begin{equation}
    \langle O\rangle_{R2}:=\frac{{\rm tr}\rho O\rho O^{\dagger}}{{\rm tr}\rho^2}.
\end{equation}

Similar to the R\'enyi-$1$ correlator, the R\'enyi-$2$ correlator also has a pure state interpretation in a doubled Hilbert space $\mathcal{H}^L\otimes\mathcal{H}^R$: it is simply an ordinary correlation function for the state $\frac{|\rho\rangle\rangle}{\sqrt{{\rm tr} \rho^2}}$. However, this doubled state, known as the Choi double, is not a purification of $\rho$, in contrast to the canonical purification $|\sqrt{\rho}\rangle\rangle$. 

Another advantage of vectorizing $\rho$ (instead of $\sqrt{\rho}$) is that time-evolution takes simple forms. For example, a continuous time Lindbladian master equation $\partial_t\rho=\mathcal{L}[\rho]$, which is a linear equation in $\rho$, take the form of a non-Hermitian Schr\"odinger equation on the Choi double state. More generally, a quantum channel can be viewed as a linear evolution on $|\rho\rangle\rangle$, which allows one to generalize various existing analytical as well as numerical techniques. Much progress has been made along this line -- a sampling of works include Refs.~\cite{Kuno2024,GuoYang2025,Shu2026,Ding2026,Sarma2026}. To study R\'enyi-$1$ or fidelity quantities in similar techniques often requires taking replica limits, making the problem much subtler.

With all the advantages of the R\'enyi-$2$ correlator, it does come with a cost: it is in general not a faithful characterization of SW-SSB. There are no bounds like Eq.~\ref{eq:R1bounds}. There are even examples~\cite{Lessa2024} that have long-range (short-range) order in fidelity, but short-range (long-range) order in R\'enyi-$2$. More importantly, most of the consequences that can be derived from fidelity long-range order, to be reviewed in later sections, do not follow from R\'enyi-$2$ long-range order. This is not surprising, as the R\'enyi-$2$ correlator reflects properties of $\rho^2$ rather than $\rho$ -- observables are weighted not by probability $p$, but by $p^2$. Thus, one should view the R\'enyi-$2$ quantity as a proxy for the intrinsic properties. Such a proxy is especially useful when $\rho$ and $\rho^2$ are not too different, which is typically the case when the system is far from phase transitions. This tradeoff between conceptual power and practical computability is quite common in information-theoretic quantities. A canonical example is the contrast between von Neumann and R\'enyi entropies: the former enjoys appealing properties, such as strong subadditivity, but computing $\rho\ln\rho$ is certainly much harder than $\rho^2$.

\subsection{Detectability and local characterization}
\label{sec:detect}

SW-SSB comes with various consequences that have operational and observable meanings, which will be reviewed in subsequent sections. However, the notion itself, defined through the various correlators discussed above, may not be easily measurable in practice. Unlike ordinary correlators $\langle O_i O_j^{\dagger}\rangle$, which can be directly measured using standard experimental techniques such as X-ray or neutron scattering, the fidelity or R\'enyi correlators require global information about the state $\rho$. Knowing the full density matrix would, of course, allow us to evaluate such quantities, but this requires full tomography, which costs an exponential amount of resources and is not practically feasible~\cite{GilyenPoremba2022}. Such difficulties in direct observation are typical for quantities that are nonlinear in the density matrix.

In the context of SW-SSB, this difficulty has been made precise in Ref.~\cite{Feng2025}, where it was argued that SW-SSB cannot be detected efficiently. Here, ``efficient'' means that the required resources -- number of copies of the state, ancilla qubits, and quantum operations -- scale polynomially with system size, ${\rm Poly}(N)$. Specifically, Ref.~\cite{Feng2025} constructed a set of ``pseudo-SWSSB'' states. A pseudo-SWSSB state is an ensemble of states $E$, in which each state $\rho\in E$ is strongly symmetric without SWSSB order (i.e., has vanishing fidelity correlator), and can be prepared efficiently. However, given $k\sim {\rm Poly}(N)$ copies of the same state $\rho$ drawn from the ensemble, the ensemble-averaged state $\mathbb{E}_E\rho^{\otimes k}$ cannot be distinguished from a truly SW-SSB state $\rho_{\rm SWSSB}^{\otimes k}$ using an efficient quantum algorithm. The existence of such pseudo-SWSSB states shows that detecting SW-SSB is, in general, difficult. Note that this is a ``worst-case-scenario'' result in complexity theory, where one assumes no prior knowledge of the state being measured. In physically relevant situations, we often do have some knowledge of the states, and such knowledge may allow us to circumvent this complexity barrier. This situation arises in the recent experiment~\cite{Experiment2026}, where the state is known to be well approximated by a free-fermion Gaussian state. Exploiting this structure, the authors were able to obtain accurate variational estimates of the fidelity and R\'enyi correlators.

There is yet another restriction in physically realistic settings: measurements are typically limited to local operators. This is certainly true in solid-state systems, and even in quantum simulators, measuring global operators such as $\otimes_i X_i$ may involve overcoming errors that grow exponentially with system size. Once we restrict ourselves to local observables, the difficulty of detecting SW-SSB becomes even more apparent: even the simplest SW-SSB state $\rho_0\sim I+X$ cannot be distinguished from a trivial maximally mixed state $\rho_{MM}\sim I$ using local observables. More formally, the difficulty in characterizing SW-SSB in terms of local observables is tied to the difficulty in defining the notion on an infinite system. Mathematically, the ``total Hilbert space'' is not well-defined on an infinite system, so quantities defined through the global density matrix $\rho$ may not always be well-defined. The reduced density matrix on a finite region $A$, however, remains well-defined and can be used to characterize the state on an infinite system~\cite{bratteli_operator_2012}.

We can gain some insight from the simple example $\rho_0\sim I+X$. Any local region $A$ appears maximally mixed: $\rho_{0A}\sim I$, with the strong symmetry explicitly broken. This explicit breaking of strong symmetry can be detected through the one-point fidelity correlator calculated using $\rho_{0A}$ (rather than the full $\rho_0$). This example suggests that we can (at least partially) detect SW-SSB by examining the level of strong-symmetry breaking in local regions. 

More generally, given the reduced density matrix $\rho_A$ on a local patch $A$, we can define the one-point local fidelity correlator for a charged operator $O_i$ acting inside $A$:
\begin{equation}
    \langle O_i\rangle_F^{(A)}:=F(\rho_A,O_i\rho_A O_i^{\dagger}).
\end{equation}
This quantity depends only on $\rho_A$, not the full density matrix $\rho$, and thus can in principle be efficiently measured if $A$ is not too large. Intuitively, the boundary of the region, $\partial A$, can in general break the strong symmetry, so we can viewed the local fidelity correlator as the fidelity correlation between a charged operator and a symmetry-breaking boundary. One can prove that in the presence of SW-SSB, the one-point local fidelity will remain nonzero as the size of region $A$ grows:
\begin{equation}
\label{eq:localdef}
    \lim_{|A|\to\infty}\langle O_i\rangle_F^{(A)}>0.
\end{equation}
This result essentially follows from the fact that fidelity is monotonically non-decreasing under partial tracing~\cite{nielsen_quantum_2010}. States satisfying the above condition are said to exhibit \textit{local SW-SSB}~\cite{Divi2026,LiuYiElse2026,Zhang2026}. Conversely, if $\langle O_i\rangle_F^{(A)}$ decays with increasing $|A|$, the state does not exhibit SW-SSB. In general Eq.~\ref{eq:localdef} is only a necessary, but not sufficient condition of SW-SSB. If we think of SW-SSB as a type of ``charge thermalization'', then Eq.~\ref{eq:localdef} is the analogue of ``local thermalization'', which does not necessarily imply global thermalization -- a familiar example is highly excited pure states satisfying the eigenstate-thermalization hypothesis (ETH)~\cite{ETHReview}. In practice, for many physically relevant examples, including those discussed in the next section, this local criterion is equivalent to the definition based on global two-point fidelity correlators. Another nice feature of the local definition is that the limit in Eq.~\eqref{eq:localdef} is guaranteed to exist on infinite systems~\cite{Divi2026,LiuYiElse2026}, even though the global density matrix itself is not defined. This provides a more satisfactory definition of SW-SSB in the thermodynamic limit.

\section{EXAMPLES}

We now review a few basic examples involving SW-SSB. 

\subsection{Thermal states}

As discussed in the previous section, SW-SSB can be understood intuitively as a form of ``charge thermalization'', reflecting the system's insensitivity to moving a charge from one point to another in space. Combined with the familiar equivalence between canonical (strong symmetry) and grand-canonical (weak symmetry) ensembles, this strongly suggests that thermal states in the canonical ensemble
\begin{equation}
    \rho_{\beta,Q}=\frac{e^{-\beta H}P_Q}{Z},
\end{equation}
with $P_Q$ projecting onto the charge-$Q$ sector, should have spontaneous breaking of the strong symmetry at any nonzero temperature ($\beta<\infty$). The symmetry-breaking pattern is strong-to-weak if ordinary SSB does not occur; otherwise, the symmetry is broken from strong to none. The simplest example is again $\rho_0\sim I+X$, which is an infinite-temperature thermal state in canonical ensemble. The statement was first derived through physical arguments in Refs.~\cite{Lessa2024,LiuRenyi1}, and more recently proved in Ref.~\cite{Zhang2026}.

\subsection{The decohered Ising model}
\label{sec:decoheredIsing}

SW-SSB can also arise in non-thermal settings, which are particularly relevant in the context of noisy quantum devices and mixed-state phases of matter. The most well-studied example is the decohered Ising model~\cite{LeeJianXu2023,ChenGrover,Lessa2024}. 

The setup is to start from a trivial paramagnetic state on a $d$-dimensional square lattice
\begin{equation}
\rho_I=\otimes_i|X_i=1\rangle\langle X_i=1|,
\end{equation}
and then subject the system to decoherence modeled by a simple quantum channel:
\begin{equation}
\label{eq:decoheredIsing}
\mathcal{E}_p=\otimes_{\langle ij\rangle}\mathcal{E}_{p,\langle ij\rangle}, \hspace{10pt} \mathcal{E}_{p,\langle ij\rangle}[\rho]=(1-p)\rho+pZ_i Z_j \rho Z_j Z_i,
\end{equation}
where $\langle ij\rangle$ denotes a nearest-neighbor bond. For $0<p<1/2$, the channel $\mathcal{E}_{p,\langle ij\rangle}$ can be interpreted as a weak measurement of $Z_i Z_j$, while the limit $p\to1/2$ corresponds to strong measurement. One may also loosely interpret $p$ as an effective decoherence time, with $p\to1/2$ corresponding to the infinite-time steady state. The channel is obviously strongly symmetric ($Z_iZ_j$ commutes with $X$).

It is interesting to consider the decohered state $\mathcal{E}_p[\rho_I]$ as a function of the decoherence strength $p$. The two limits are simple: at $p=0$ the state is pure and no SSB in any sense; at $p=1/2$ we have $\mathcal{E}_{p=1/2}[\rho_I]=(I+X)/2^N$, which has SW-SSB order. Something interesting must happen in between.

The phase transition can be studied by mapping the problem to classical statistical-mechanical models, in the same space dimension. Details of this mapping can be found in Refs.~\cite{zhu2022nishimoris,Lee2022,LeeJianXu2023,ChenGrover,Lessa2024}. The decoherence strength becomes an effective temperature, with the $p\to0$ and $p\to1/2$ limits as high and low temperature limits, respectively. Thus in $1d$ there is no transition until the zero-temperature $p\to1/2$ limit. In $2d$ there is a continuous transition at some intermediate $p_c^{(n)}$, which in general will depend on the R\'enyi index $n$.  For $n=2$ (relevant for R\'enyi-$2$ correlators) the model becomes the standard Ising model with $p_c^{(2)}\approx0.178$. For the intrinsic transition relevant for R\'enyi-$1$ and fidelity correlators ($n=1$), the model is the Kramers-Wannier dual of celebrated random-bond Ising model (RBIM) along the Nishimori line~\cite{Nishimori1980}, with a continuous transition at $p_c\approx 0.109$. The fidelity or R\'eny-$1$ correlator maps to the $1/2$-moment of the disorder operator in the Nishimori RBIM, which at $p_c$ decays as a power-law $1/|x-y|^{\eta}$. The current numerical estimates give $\eta\approx0.16-0.17$.~\cite{MerzChalker2003,wang_analog_2024} Near the transition, the fidelity correlator decays exponentially, with diverging correlation length $\xi\sim |p-p_c|^{-\nu}$, and from existing numerics $\nu\approx 1.33-1.52$.~\cite{merz2002rbim,RBIM1,RBIM2,RBIM3,wang_analog_2024}

The decohered Ising model is closely related to a well-studied problem in quantum information: the decodability transition of the toric code under $X$ or $Z$ dephasing\cite{TQM}. The Nishimori transition in this context has long been understood~\cite{TQM,WangPreskill2003}, and recent works have reinterpreted it (along with its higher R\'enyi generalizations) as a transition between distinct mixed-state phases~\cite{FanBaoMixedTopo,BaoFandouble,ChenGrover2}. The connection between the two problems follows from the standard Kramers–Wannier duality. Recall that at zero temperature, tuning the Hamiltonian drives the toric code out its topological order via condensation (Higgsing) of the $\mathbb{Z}_2$ gauge charge\cite{KogutReview}. The decodability transition has a similar interpretation: since gauge symmetry is necessarily strong in mixed states (by definition, only gauge-invariant states are physical), the decodability transition can be viewed as a ``strong-to-weak Higgsing'' of the $\mathbb{Z}_2$ gauge symmetry.

\subsection{Experimental realization in dephased Fermi gas}
\label{sec:experiment}

Recently, evidence of SW-SSB in terms of fidelity/R\'eny correlators has been observed in cold atoms~\cite{Experiment2026}. Specifically, one starts from an initial state $\rho_0$, and take ``snapshots'' of the atoms by simultaneously measuring the particle occupation number on each lattice site. Averaging over the snapshots is equivalent to maximal dephasing in the density operator. If $\rho_0$ is an insulating ground state, the charge is localized on each site and unaffected by the measurement, hence one does not expect SW-SSB. If $\rho$ is metallic, described at zero temperature by a Fermi surface, then the charge is already fluctuating nontrivially in the ground state, and the maximal dephasing will induce SW-SSB order~\cite{Experiment2026}. At finite temperature, SW-SSB order exists for both states, but the changes in the fidelity/R\'enyi correlators under dephasing are still expected to be quite different in a metal and an insulator. By varying the optical lattice potential, a metal-insulator transition is first induced in the initial state, and the change in the behavior of fidelity/R\'enyi correlator has been observed after taking snapshots~\cite{Experiment2026}. This marks the first experimental evidence of the SW-SSB order.

\subsection{Other examples}
\label{sec:otherexamples}

Many more examples of SW-SSB and associated phase transitions have been studied in recent literature. The physical setup is quite diverse, including decohered ground state phase transitions~\cite{GuoYangYu}, SW-SSB transitions in Lindbladian steady states~\cite{Lin2025,Zhang2026} or in the presence of weak post-selection~\cite{Knap2025},  and decohered highly entangled system such as the Sachdev-Ye-Kitaev (SYK) model~\cite{SYK1,SYK2}. The subject remains under active exploration.

\section{INFORMATION-THEORETIC CONSEQUENCES} 

\subsection{Long-range conditional mutual information}
\label{sec:CMI}

In what sense is the fidelity correlator Eq.~\ref{eq:FC} a ``correlator''? What kind of ``correlation'' does it capture? To answer this question, first recall that the usual correlation function $\langle O_iO^{\dagger}_j\rangle$ captures how much we know about $O_j$ if we measure $O_i$. This correlation is captured by an information-theoretic quantity known as the mutual information.

\begin{figure}[h]
\includegraphics[width=1.5in]{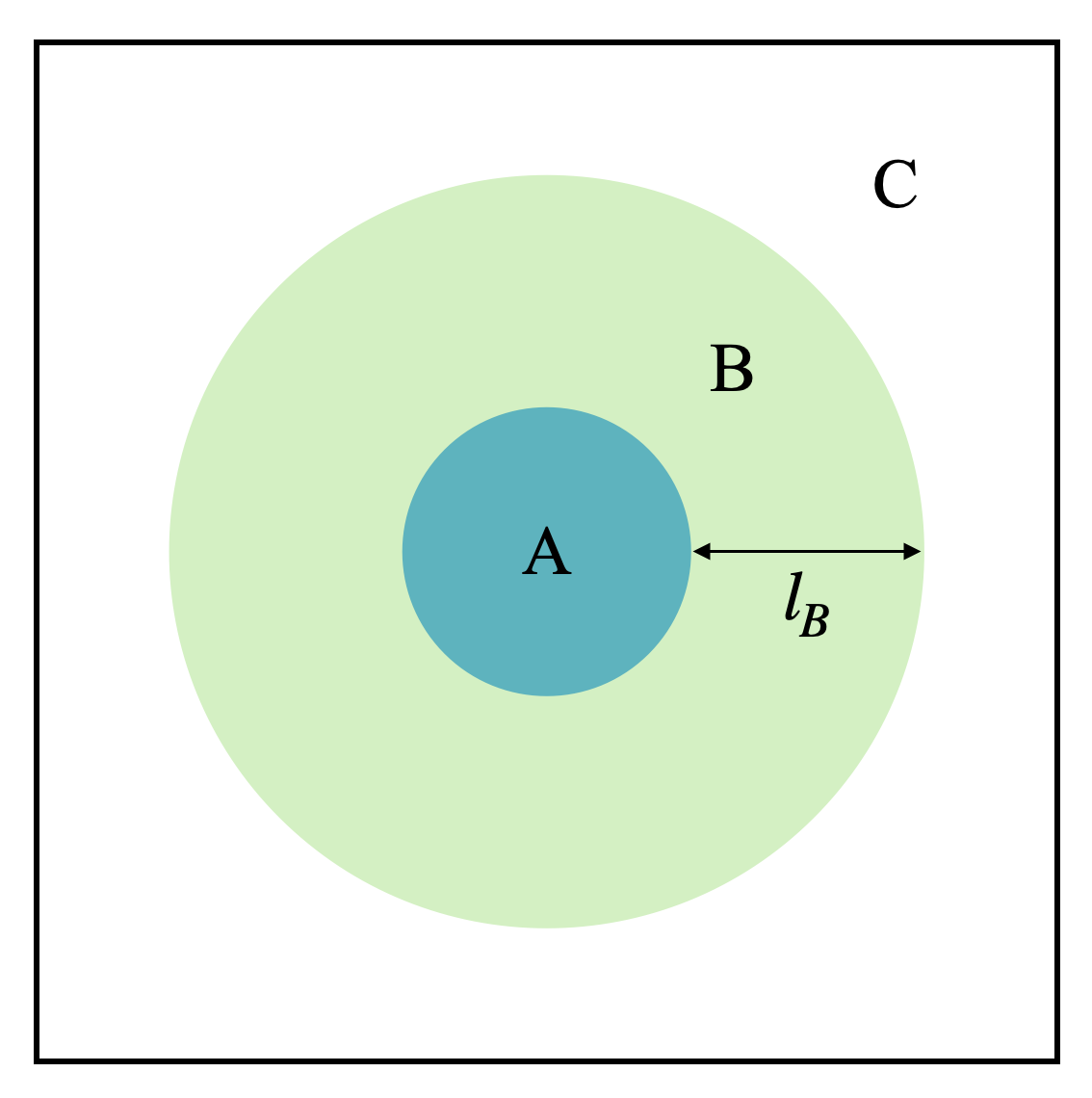}
\caption{Configuration of the tripartition: $A$ is a disc, $B$ is a buffer region around $A$, and $C$ is the rest of the system. We are mainly interested in the limit of large $l_B$.}
\label{fig:ABC}
\end{figure}

Let us partition the system into three parts, $A$, $B$, and $C$, as shown in \textbf{Figure~\ref{fig:ABC}}. Regions $A$ and $C$ are separated by the buffer region $B$, whose width $l_B$ will be taken to be large in later discussions. The mutual information (MI) between $A$ and $C$ is
\begin{equation}
    I(A,C)=S_A+S_C-S_{AC},
\end{equation}
where $S_{R}$ denotes the von Neumann entropy of the reduced density matrix on region $R$. This quantity tells us how much information we can gain about region $A$ ($C$) by knowing about region $C$ ($A$). It is well known~\cite{MICorr2008} that the MI upper-bounds connected two-point correlation functions between region $A$ and $C$. This means that if $I(A,C)\sim e^{-l_B/\xi_c}$, then any two-point correlation function between $A$ and $C$ should decay exponentially -- the $\xi_c$ from the MI thus gives an operator-independent way to define correlation length. Conversely, if ordinary SSB occurs and the connected two-point function of some charged operator do not decay, then $I(A,C)$ will also stay nonzero as $l_B\to\infty$.

Recently, the importance of another quantity, the conditional mutual information (CMI), has been increasingly appreciated in the context of many-body physics~\cite{sang2024stability}. In the same tripartite configuration as \textbf{Figure~\ref{fig:ABC}}, the CMI between region $A$ and $C$, conditioned on $B$, is
\begin{eqnarray}
    I(A,C|B)&=&S_{BC}+S_{AB}-S_B-S_{ABC} \nonumber \\
    &=&I(A,BC)-I(A,B).
\end{eqnarray}
The second line makes the meaning of CMI apparent: it quantifies the correlation between $A$ and $C$ that is not mediated through $B$. CMI has always been an important concept in (quantum) information theory~\cite{nielsen_quantum_2010}. In condensed matter physics it appears as the topological entanglement entropy in the Levin-Wen prescription~\cite{LevinWen}. For us two facts about CMI are particularly important:
\begin{enumerate}
    \item Strong sub-additivity (SSA): $I(A,C|B)\geq0$. 
    \item Approximate local recoverability: consider a quantum channel $\mathcal{E}_A$ acting on region $A$ (it can be as violent as tracing out region $A$), we ask if the original state $\rho$ can be ``locally recovered'' via another quantum channel $\tilde{\mathcal{E}}_{AB}$ acting on the enlarged region $AB$ (importantly, not on $C$). It turns out such a recovery exists approximately if $I(A,C|B)$ is small~\cite{Recoverability}. More precisely, there is an optimal Petz recovery map~\cite{Petz1986} $\tilde{\mathcal{E}}_{AB}$, which depends on both $\rho$ and $\mathcal{E}_A$, such that
    \begin{equation}
    \label{eq:recover}
        F(\rho,\tilde{\mathcal{E}}\cdot\mathcal{E}[\rho])\geq 2^{-I(A,C|B)/2}.
    \end{equation}
    So for $I(A,C|B)=0$ (known as a quantum Markov chain state), the recovery can be exact. For small $I(A,C|B)$ (known as an approximate quantum Markov chain), the recovery can be done to a good approximation. The intuition is that with small CMI, the correlations between $A$ and $C$ are mostly mediated through $B$, so even if we lose $A$, being able to act on $AB$ should be enough to recover most of the correlations in the system. 
\end{enumerate}

Note that for a pure state, $S_{ABC}=0$, and it is easy to see that the MI and CMI defined above are identical. Thus, CMI as a distinct measure of correlation is a feature of mixed states. Recall that when MI decays exponentially in $l_B$, the corresponding length scale $\xi_c$ is interpreted as the correlation length. Similarly, when CMI decays exponentially, $I(A,C|B)\sim e^{-l_B/\xi_M}$, the corresponding length scale is called the ``Markov length''~\cite{sang2024stability}. The significance of a finite $\xi_M$ is that any local channel on $A$ can be recovered to a very good approximation if the recovery channel can act on $AB$ with $l_B\gg \xi_M$.

The correlation probed by the CMI is very different from that captured by the MI. This is best illustrated by thermal Gibbs states at finite temperature, in the ``grand-canonical'' ensemble so the global symmetry is only weak. For a local Hamiltonian at finite temperature, it has been proven that the CMI always decays exponentially in such states. More precisely, Ref.~\cite{Chen2025} proved that for a local region $A$ ($|A|\sim O(1)$), the CMI is bounded by $|C|e^{-l_B/\xi_M}$. The bound can be improved in one dimension~\cite{Kato2016,Kuwahara2025}, where the CMI is bounded by $e^{-l_B/\xi_M}$ without assuming $|A|\sim O(1)$. In particular, even across a finite-temperature phase transition, where the MI typically develops a singularity, the CMI is generally smooth and does not detect the transition.

The simplest example that does have long-range CMI (meaning $I(A,C|B)$ does not decay with $l_B$), but with short-range MI is the maximally mixed strongly symmetric state $\rho_0\sim I+X$. A straightforward calculation shows that $I(A,C|B)=\log 2$ (independent of $l_B$) and $I(A,C)=0$. We can also understand this long-range CMI from recoverability: if we trace out a region (even a single qubit), the state becomes $\rho'\sim I$, and the information of the global $\mathbb{Z}_2$ charge is lost. There is no way to act with a local quantum channel to recover the global constraint of the $\mathbb{Z}_2$ charge. Based on Eq.~\ref{eq:recover} this means that the CMI should be non-zero.

More generally, it has been proved~\cite{Lessa2024} that if a strongly symmetric state $\rho$ exhibits SW-SSB order, then CMI should be long-ranged, i.e. $I(A,C|B)>0$ as $l_B\to\infty$. This is true even if the SW-SSB order holds only at the local level~\cite{Divi2026,Zhang2026} as in Eq.~\eqref{eq:localdef}. The argument roughly goes as follows: starting from the nonzero local fidelity correlator
\begin{equation}
\label{eq:CMIproof1}
    F(\rho_{AB},O_i\rho_{AB} O_i^{\dagger})>0,
\end{equation}
where $O_{i}$ denotes a local charged operator deep inside region $A$, and the buffer region $B$ is arbitrarily large but still smaller than the whole system. Crucially, the fidelity $F(\rho,\sigma)$ satisfies the \textit{data processing inequality} (DPI)~\cite{nielsen_quantum_2010}, meaning that it is non-decreasing under any quantum channel $F(\mathcal{E}[\rho],\mathcal{E}[\sigma])\geq F(\rho,\sigma)$. We then act on $\rho_{AB}$ with an arbitrary channel in region $BC$, call it $\tilde{\mathcal{E}}_{B\to BC}$, and ask if we can possibly recover $\rho$. $\tilde{\mathcal{E}}_{AB}$ commutes with $O_C$ since they do not overlap. So again by DPI, Eq.~\ref{eq:CMIproof1} implies that
\begin{equation}
    F(\tilde{\mathcal{E}}_{B\to BC}[\rho_{AB}],O_i\tilde{\mathcal{E}}_{B\to BC}[\rho_{AB}]O^{\dagger}_i)>0.
\end{equation}
This means that the attempted recovery will always lead to a state $\tilde{\mathcal{E}}_{B\to BC}[\rho_{AB}]$ whose one-point fidelity correlator is non-zero, so the strong symmetry cannot be recovered. Therefore the original state $\rho$ cannot be recovered (even approximately), and by Eq.~\ref{eq:recover} the original state must have long-range CMI.

\begin{table}[h]
\tabcolsep7.5pt
\caption{Different SSB patterns, their representative states, and information-theoretic correlations.}
\label{Tab:correlation}
\begin{center}
\begin{tabular}{@{}l|c|c|c@{}}
\hline
SSB type & Representative state& Long-range MI? & Long-range CMI?\\
\hline
Strong-to-weak & $\rho_0=\frac{I+X}{2^N}$ & No &Yes\\
\hline
Weak-to-none & $\rho=\frac{1}{2}(|\uparrow\uparrow..\rangle\langle\uparrow\uparrow..|+|\downarrow\downarrow..\rangle\langle\downarrow\downarrow..|)$ &Yes &No\\
\hline
Strong-to-none  &$|\Psi\rangle=\frac{1}{\sqrt{2}}(|\uparrow\uparrow..\rangle+|\downarrow\downarrow..\rangle)$ &Yes &Yes\\
\hline
\end{tabular}
\end{center}
\end{table}

It is quite illuminating to put all types of SSB together and examine their information-theoretic characterization (see \textbf{Table~\ref{Tab:correlation}}). The lesson is that the breaking of strong symmetry is signaled by long-range CMI, while the breaking of weak symmetry is signaled by long-range MI. The cat state has strong-to-none SSB, so naturally it has both long-range MI and CMI.

\subsection{Phases of matter from local reversibility}
\label{sec:LR}

Traditional SSB is a property of a phase, not just a state. We expect this to also be true for SW-SSB, but we have thus far not answered the seemingly simple question: what exactly is a ``phase'' in the context of mixed quantum states? This is by no means a simple question. Over the years many different definitions of mixed-state phases of matter have been proposed~\cite{hastings2011nonzero,coser2019classification,sang2023mixed,de2022symmetry,ma2023average,StableOpen2024,lessa2025higher,Sang2025LR}. The relevance of each definition may depend on the physical situation, as well as the type of questions we are interested in. 

Here we will briefly review one particular definition of mixed-state phases based on ``locally reversible channels''~\cite{Sang2025LR}. This definition has the advantage of being able to distinguish many different phases of matter, especially topological phases of matter including both SPT and intrinsic topological orders. It also distinguishes phases with and without long-range CMI.

\begin{figure}[h]
\includegraphics[width=3.5in]{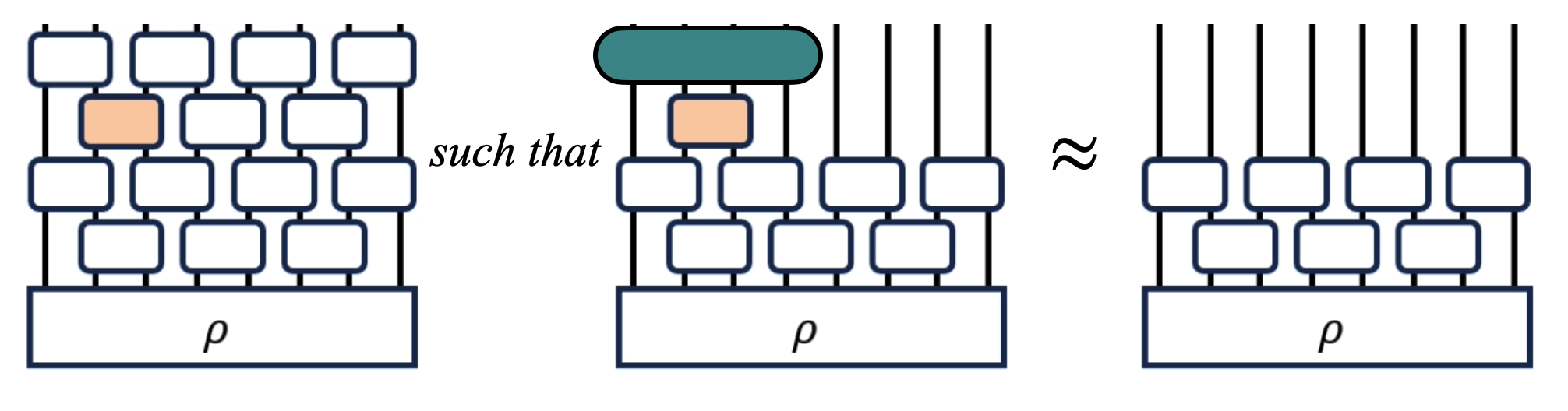}
\caption{Left: a finite-depth channel, where each gate represents a quantum channel acting on a local region, and ``depth'' refers to the number of layers in the circuit; Right: local reversibility, where the effect of each gate can be reversed by another local gate.}
\label{fig:LRChannel}
\end{figure}

Consider two states $\rho_1$ and $\rho_2$. We say that the two states belong to the same phase if and only if the following conditions are all satisfied:
\begin{enumerate}
    \item There exists a finite-depth (or more accurately, ${\rm PolyLog}(N)$ depth) circuit of quantum channels $\mathcal{E}$ (see \textbf{Figure~\ref{fig:LRChannel}}), such that $\mathcal{E}[\rho_1]\approx\rho_2$. Here and below, the $\approx$ sign means equal up to an error terms that can be made polynomially small in system size. We can think of this channel circuit as a discretized Lindbladian evolution over some finite-time $\tau$: $\mathcal{T}e^{\int_0^\tau\mathcal{L}(t)dt}[\rho_1]\approx\rho_2$.
    \item A similar reverse finite-depth channel $\tilde{\mathcal{E}}$ also exists: $\tilde{\mathcal{E}}[\rho_2]\approx\rho_1$. In other words, the effect of $\mathcal{E}$ on the state $\rho_1$ can be reversed through another finite-depth channel. 
    \item Furthermore, the above channels are also \textit{locally reversible}, in the sense that if we only evolve $\rho_1$ using a subset of gates of $\mathcal{E}_{FD}$ inside a finite region $R$ (denote the restricted channel as $\mathcal{E}_R$), its effect can also be reversed by $\tilde{\mathcal{E}}$ restricted to $R^+$ (region $R$ slightly enlarged by a finite length) (see \textbf{Figure~\ref{fig:LRChannel}}). 
\end{enumerate}

The motivation for the above (somewhat involved) definition is to mimic the definition of pure state phases, based on adiabatic evolution, or equivalently finite-depth unitary circuit~\cite{ChenGuWenLRE}. A unitary gate is automatically reversible since $UU^{\dagger}=I$. However, this is not true for quantum channels, so we have to impose the (local) reversibility condition separately. 

The behavior of CMI plays a crucial role in distinguishing different phases of matter. It has been shown that if $\rho_1$ has finite Markov length (exponentially decaying CMI), and there is a finite-time Lindbladian evolution that takes $\rho_1$ to $\rho_2$ (condition $1$ above), then $\rho_1$ and $\rho_2$ belong to the same phase if (Ref.~\cite{sang2024stability}) and only if (Ref.~\cite{Yi2026}) the Markov length remains finite throughout the entire time-evolution. In other words, if $\rho_{1,2}$ belong to different phases, then the Markov length has to diverge at some intermediate time. In our context, since SW-SSB implies diverging Markov length, we can conclude that states with SW-SSB and trivial states (such as the maximally mixed state $\rho\sim I$) belong to different mixed-state phases, even when symmetry is not taken into account. For the decohered Ising model reviewed in Sec~\ref{sec:decoheredIsing}, by examining the behavior of the Markov length, one can show that there are indeed only two phases (the symmetric phase and SW-SSB) separated by the transition at $p_c\approx0.109$.~\cite{sang2024stability,Lessa2024}

Defining phases in terms of local recoverability has proven useful for studying topological phases in mixed states. Some key lessons from the past few years include:
\begin{enumerate}
    \item The mixed-state analogue of ``gapped phases'' -- the simplest class of pure-state phases -- should be characterized by exponentially decaying MI and CMI. Equivalently, both the correlation length $\xi_c$ and the Markov length $\xi_M$ are finite. While the notion of a finite correlation length is familiar, the meaning of a finite Markov length deserves clarification. In a pure-state gapped phase, local operators create only trivial excitations, whereas nontrivial excitations (such as anyons) require nonlocal operators like string operators. The mixed-state analogue is that a finite Markov length ensures that any local operation can be approximately reversed by another local operation, and thus does not affect the global properties of the state.
    \item With the above definition of ``gapped'' mixed states, one can show~\cite{LessaChengWang} that a gapped phase cannot occur in $1d$ systems with anomalous symmetry, such as those realized on the boundary of a $2d$ SPT phase. The anomaly may involve either only strong symmetry or a combination of strong and weak symmetries~\cite{ma2023average,Maetal2025} -- so the result essentially requires either the strong or weak symmetry to be close to spontaneous breaking~\cite{XuJian2025,YouOshikawa}. This result closely parallels the familiar statement for pure-state SPT phases, where the boundary necessarily remains gapless. In fact, SW-SSB, which is not gapped in our definition due to diverging Markov length, may loosely be viewed as a special case of anomalous symmetry if we have $O_{ij}\rho O_{ij}^{\dagger}\approx\rho$, in which case $O_{ij}$ can be viewed as a $d$-form weak symmetry ($d$ being the space dimension), and this $d$-form symmetry has a mutual anomaly with the strong $0$-form symmetry. 
    \item In $d>1$, gapped states -- defined by finite $\xi_c$ and $\xi_M$ -- can exist in the presence of anomalous \textit{higher-form symmetries}~\cite{gaiotto_generalized_2014}. In the mixed-state setting, these correspond precisely to \textit{topological orders}, with the associated anomalies encoding anyon braiding phases and their higher-dimensional generalizations~\cite{GeneralizedStatistics}. Compared to the pure-state case, a key difference is that some of these anomalous symmetries may be weak. 

    A central result is that anomalous symmetries are stable under locally reversible channel deformations~\cite{Sang2025LR}: even if the deformation breaks the symmetry microscopically, it persists in a renormalized form, paralleling the behavior of pure-state topological order. This leads to a unified framework for mixed-state topological phases in terms of strong and weak higher-form symmetries and their anomalies. The weak symmetry that has a mutual anomaly with the strong symmetry can also be viewed as the order parameter of strong-to-weak breaking~\cite{Zhang2025}, further unifying the picture. In this discussion, the requirement of local reversibility is crucial. If one only imposes global reversibility (also known as ``two-way channel connectivity''~\cite{coser2019classification}, which include our conditions $1$ and $2$, but not $3$), weak symmetries are generally unstable under deformations and therefore do not contribute to the universal topological data. These structures also arise elegantly within the ``topological bootstrap'' program~\cite{MixedBootstrap2025}, which assumes exact vanishing (instead of exponentially decay) of MI and CMI, together with a small set of additional axioms.
    
\end{enumerate}

Inspired by the developments on quantum mixed states, the long-range behavior of CMI has also been studied in various classical non-equilibrium systems and surprising results have been found -- see Refs.~\cite{Lloyd2025,ChenGrover2025}.

\section{DYNAMICAL CONSEQUENCES}

\subsection{Stability theorem}
\label{sec:stability}

We now review the ``Stability theorem'' in Ref.~\cite{Lessa2024} and its implications. The theorem states that, if $\rho$ is a strongly symmetric state exhibiting SW-SSB, and $\mathcal{E}$ is a strongly symmetric finite-depth quantum channel, then $\mathcal{E}[\rho]$ will also exhibit SW-SSB. In other words, SW-SSB is stable under symmetric local evolution over finite time (finite depth). Unlike the discussion in Sec.~\ref{sec:LR}, the channels considered here need not be locally reversible, but have to be strongly symmetric.

If we think of SW-SSB as a kind of ``charge thermalization'', then very loosely speaking the above result has the flavor of an ``arrow of time'': it is often easy for a state to enter SW-SSB, but once SW-SSB is established, it is hard to exit.

To provide some intuition, we give a simplified proof of the stability theorem in the special case when the charged operator $O_i$ commutes with the channel $\mathcal{E}$ -- an example satisfying this assumption is the decohered Ising model in Sec.~\ref{sec:decoheredIsing} with $O_i=Z_i$. Then we have
\begin{equation}
    F(\mathcal{E}[\rho],O_{ij}\mathcal{E}[\rho]O_{ij}^{\dagger})=F(\mathcal{E}[\rho],\mathcal{E}[O_{ij}\rho O_{ij}^{\dagger}])\geq F(\rho,O_{ij}\rho O_{ij}^{\dagger})>0,
\end{equation}
where the first equality comes from assuming $O$ and $\mathcal{E}$ commute, the second inequality is the monotonicity of fidelity under quantum channels~\cite{nielsen_quantum_2010}, and the last inequality is just the assumption that $\rho$ has SW-SSB. In more general situations where $O_i$ may not commute with $\mathcal{E}$, Ref.~\cite{Lessa2024} proved a weaker bound for some local operator $O_i'$ with the same symmetry charge
\begin{equation}
\label{eq:stabilitybound}
    F(\mathcal{E}[\rho],O'_{ij}\mathcal{E}[\rho](O'_{ij})^{\dagger})\gtrsim  e^{-{\rm Poly(D)}} F(\rho,O_{ij}\rho O_{ij}^{\dagger}),
\end{equation}
where $D$ is the depth of the channel $\mathcal{E}$. For finite $D$ the conclusion does not change.

The stability theorem also holds for local SW-SSB, defined through Eq.~\eqref{eq:localdef} One consequence is that even in infinite systems, where it becomes subtle to define notions such as strong symmetry and global CMI, the stability theorem guarantees that SW-SSB is a universal property of mixed-state quantum phases.

It is helpful to revisit the decohered Ising model in Sec.~\ref{sec:decoheredIsing}. The noise channel Eq.~\ref{eq:decoheredIsing} is a strongly symmetric channel with finite depth. This example shows that one can start from a fully symmetric state (the product paramagnetic state $|+++...\rangle$) and reach SW-SSB through a finite-depth symmetric channel (the noise channel at $p>p_c\approx0.109$). The stability theorem tells us that the reverse cannot be done: starting from $\mathcal{E}_{p>p_c}[|++...\rangle\langle++...|]$, there is no symmetric finite-depth channel that can bring us back to a state at $p<p_c$. 

In the decohered Ising example, it is possible to construct, using local Petz recovery maps~\cite{sang2024stability,Lessa2024}, a strongly symmetric shallow channel circuit that takes a state from $p>p_c$ to another state at $p'$ satisfying $p_c<p'<p$. Namely, the degree of SW-SSB can be somewhat reduced -- just not to zero. In particular, one can start from $p>p_c^{(2)}\approx 0.178$, the transition point for the R\'enyi-$2$ correlator, and move to $p'<p_c^{(2)}$ as long as $p'>p_c$. This example explicitly shows that the stability theorem applies only to SW-SSB defined through R\'enyi-$1$ quantities, such as the fidelity or R\'enyi-$1$ correlator, but not to the R\'enyi-$2$ correlator.

\subsection{Steady-state SW-SSB}
\label{sec:steady}

Since strong symmetry is generically spontaneously broken in thermal states, either to weak symmetry or to no symmetry at all, it is natural to expect that dissipative dynamics respecting the strong symmetry will likewise generically produce steady states that spontaneously break it. This expectation is confirmed in the study of typical Lindbladian and quantum-channel models, whose steady states exhibit SW-SSB~\cite{Knap2025}. For discrete symmetries, Ref.~\cite{LuHolography2025} further provides a topological-holographic argument suggesting that SW-SSB is a generic phenomenon. 

Refs.~\cite{Lin2025,Zhang2026} demonstrated that SW-SSB can be avoided in the steady states of certain local Lindbladians, where the dynamics is essentially described by the classical reaction-diffusion theory. However, doing so requires fine-tuning the dynamics and having a mixing time (not the inverse spectral gap) that diverges with the system size $L$. This naturally suggests the conjecture that any local channel dynamics whose steady states avoid SW-SSB must necessarily be slow-mixing. 

This conjecture is also supported by the stability theorem of Sec.~\ref{sec:stability}. Consider a strongly symmetric local Lindbladian, or more generally a low-depth quantum channel, for which every initial state rapidly converges to a steady state. Starting from an initial state that exhibits SW-SSB order, one is naturally led to expect that at least some of the resulting steady states must retain SW-SSB order.  This does not constitute a rigorous proof, as the bound given by the stability theorem, Eq.~\eqref{eq:stabilitybound}, decays too fast with the circuit depth $D$. The situation is better in one dimension: with some additional mild assumptions, one can argue that steady states without SW-SSB order must be slow-mixing due to the non-vanishing of \textit{disorder parameters}~\cite{Ma2024SPTdoubled}.

\subsection{Continuous symmetry and hydrodynamics}
\label{sec:hydro}

Continuous symmetry comes with additional consequences even in conventional SSB, including the Goldstone modes and the Hohenberg-Mermin-Wagner (HMW) constraints. It is natural to examine similar consequences of continuous symmetry in SW-SSB. This has been an active topic in recent years, and deep connections have been made between $U(1)$ SW-SSB and the emergence of hydrodynamics. We summarize some results below:
\begin{enumerate}
    \item If we start from a fully symmetric state (e.g., a pure product state) and evolve under a strongly $U(1)$-symmetric Lindbladian, there could be a transition to SW-SSB at some finite time $\tau$ (similar to $p<1/2$ in the decohered Ising model, Eq.~\ref{eq:decoheredIsing}) only in spatial dimensions $d\geq 2$. This is the SW-SSB analogue of the HMW theorem. For $d=2$, the transition is BKT-like, and the quasi-ordered phase has R\'enyi-$1$ and $2$ correlators that decay as power laws.~\cite{U1SWSSB1,U1Hydro,Vijay2025BKT} This can be viewed as the $U(1)$ analogue of the decohered Ising model in Sec.~\ref{sec:decoheredIsing}.
    \item As $\tau\to\infty$, the system typically approaches a steady state with SW-SSB in any dimension. The Goldstone mode gives rise to a diffusive mode in an emergent hydrodynamic description~\cite{OgunnaikeFeldmeierLee2023,Moudgalya_2024,Watanabe2024,GuWangWang2025,HydroHuang}. Following Ref.~\cite{HydroHuang}, the effective Lagrangian can be written in terms of the Goldstone field $\varphi$ and its canonical conjugate $n$ (the charge density):
    \begin{equation}
        \mathcal{L}[n,\varphi]=n\partial_t\varphi+i\sigma(n)\nabla\varphi\cdot\nabla(\varphi+i\mu(n)),
    \end{equation}
    where $\mu(n)$ is the chemical potential. The charge susceptibility is given by $\chi=\partial n/\partial\mu\geq0$. Varying $\varphi$ leads to the standard diffusion equation for density $n$.  
    The absence of HMW constraint comes from the quadratic dispersion of the Goldstone (diffusion) mode required by detail balance. Furthermore, the onset of SW-SSB and emergent hydrodynamics can also be reflected in the behavior of equal-time charge fluctuation~\cite{HydroHuang,Lee2026}.
    \item In $d\geq2$, the SW-SSB transition time $\tau_c$ can be viewed as a sharp definition of the onset point of emergent hydrodynamics. The situation is more subtle in $1d$: neither SW-SSB nor a standard hydrodynamic description emerges at finite time, yet diffusive modes can still arise in certain models, obscuring the relation between diffusion and SW-SSB. A possible clue is that, in $1d$, the characteristic length scale associated with SW-SSB (e.g., the Markov length) was found to grow ballistically, $\sim t$, which is parametrically faster than diffusive transport, $\sim \sqrt{t}$~\cite{U1Hydro}. This suggests that the system effectively exhibits SW-SSB under the $O(t)$ length scale, which leads to the diffusive mode at much shorter $O(\sqrt{t})$ length scale.
\end{enumerate}

While the theory of $U(1)$ SW-SSB appears to be largely classical, the situation is very different for continuous non-Abelian groups such as $SU(2)$. Consider a strongly $SU(2)$-symmetric channel that is unital (for example, a Lindbladian with Hermitian jump operators). The steady state is the maximally mixed state within the singlet sector. This state exhibits SW-SSB of $SU(2)$, but in contrast to previous examples, it is also highly entangled~\cite{Livine2005,Sahu2024SU2,Li2025SU2}: entanglement measures (formation, distillation, and negativity) scale as $\log(N)$. This clearly goes beyond classical hydrodynamics, and a systematic and universal understanding of the interplay between SW-SSB and many-body entanglement is still lacking.

\section{DISCUSSIONS}

The theory of SW-SSB has emerged as a powerful framework that is able to connect many ideas in a unified fashion: symmetry breaking, thermalization, Edwards–Anderson order, topological order, open-system dynamics, and (quantum) information concepts such as conditional mutual information and recoverability. We have reviewed some key developments in this work. We conclude by outlining a few future and ongoing directions.

\begin{enumerate}

\item \textit{Entanglement}: the prototypical state we discussed, $\rho_0\sim I+X$, is completely classical and has no quantum entanglement. However, more exotic states exist that exhibit both SW-SSB and nontrivial quantum entanglement. This will happen when the underlying strong symmetry has some nontrivial feature: examples include symmetries with 't Hooft anomaly~\cite{LessaChengWang,WangLi2025}, non-Abelian (especially Lie group) symmetries~\cite{Sahu2024SU2,Li2025SU2,Sahu2026,Negari2025}, and even lattice translation symmetry~\cite{Translation1,Translation2}. A deeper and more systematic understanding of many-body entanglement in mixed-state phases remains a fascinating direction for future work.

\item \textit{Generalized symmetry}: instead of ordinary global symmetry, one can discuss higher-form symmetries and their spontaneous breaking, either from strong to weak or from strong to none. This leads to the theory of mixed-state topological orders~\cite{Wang2021mixed,FanBaoMixedTopo,BaoFandouble,ChenGrover2,WangWuWang,RamanjitAbhinav,EllisonCheng,LiLeeYoshida2025,Grover2025,lessa2025higher,HsinKobayashiRem2025,Sang2025LR,MixedBootstrap2025,SongZhang2025}. On a related front, one can incorporate the powerful framework of topological holography (SymTFT) into the discussion of mixed-state phases -- see Refs.~\cite{Sun2025,LuHolography2025,SymTFT1,SymTFT2,SymTFT3,Vijay2025Holo} for some recent works.

\item \textit{Learning and inference}: one can phrase SW-SSB as the loss of our ability to efficiently learn about the system, especially on inferring the global charge. This intuition has been made precise in recent works, especially on the connection to ``charge-sharpening''~\cite{Sharpening} in monitored quantum circuits in  Refs~\cite{Learnability2026,U1Hydro,Zerba2025}, and on the unlearnability of SW-SSB through classical machine learning in Ref.~\cite{Unlearnable}. It will be interesting to explore further connections in this direction.

\item \textit{Phase transition}: as discussed in Sec.~\ref{sec:otherexamples}, phase transitions involving SW-SSB, in various different contexts, have been actively explored recently. A unifying picture is still under development. In light of the discussion in Sec.~\ref{sec:steady}, more can be explored on phases that avoid SW-SSB -- similar to various interesting phases that avoid thermalization.

\item \textit{Insights on pure states}: having developed new organizing principles for many-body physics in mixed states, one may ask whether these ideas can be brought back to the study of pure states, such as ground states of local Hamiltonians. In particular, can they provide new perspectives on the interplay between symmetry and entanglement in nontrivial ground-state phases?

\item \textit{Detection}: The recent experimental observation of SW-SSB in a cold Fermi gas platform~\cite{Experiment2026} marks an important step toward the experimental study of mixed-state quantum phases. It demonstrates that SW-SSB -- and more broadly, phases characterized by information-theoretic quantities beyond conventional correlation functions -- can be probed on modern quantum platforms. Looking ahead, it will be exciting both to identify new experimentally accessible signatures and to develop scalable measurement protocols. The local formulation of SW-SSB discussed in Sec.~\ref{sec:detect} may provide a useful route toward this goal~\cite{Divi2026,LiuYiElse2026,Zhang2026}.

\end{enumerate}

\section*{DISCLOSURE STATEMENT}
The author is not aware of any affiliations, memberships, funding, or financial holdings that
might be perceived as affecting the objectivity of this review. 

\section*{ACKNOWLEDGMENTS}
My understanding of this subject has largely been shaped by fruitful collaborations and/or discussions with Zhen Bi, Meng Cheng, Francisco Divi, Dominic Else, Amelie Gilardi, Tarun Grover, Timothy Hsieh, Xiaoyang Huang, Leonardo Lessa, Ruizhi Liu, Peter Lu, Ruochen Ma, Amin Moharramipour, Roger Mong, Subhayan Sahu, Shengqi Sang, Alex Turzillo, Cenke Xu, Yizhi You, Jinmin Yi, Carolyn Zhang, Jianhao Zhang and Yijian Zou. Research at Perimeter Institute is supported in part by the Government of Canada through the Department of Innovation, Science and Industry Canada and by the Province of Ontario through the Ministry of Colleges and Universities. This manuscript benefited from language improvements assisted by ChatGPT 5.

\bibliographystyle{ar-style4}

\bibliography{main}

\begin{thebibliography}{122}
\expandafter\ifx\csname natexlab\endcsname\relax\def\natexlab#1{#1}\fi

\bibitem{LandauLifshitz}
Landau LD, Lifshitz EM. 1980.
{Statistical Physics, Part 1}.
vol.~5 of \textit{Course of Theoretical Physics}.
Oxford: Butterworth-Heinemann

\bibitem{McGreevy2023}
{McGreevy} J. 2023.
\textit{Annual Review of Condensed Matter Physics} 14:57--82

\bibitem{GarrattChalker2021}
Garratt SJ, Chalker JT. 2021.
\textit{Phys. Rev. Lett.} 127(2):026802

\bibitem{LeeJianXu2023}
Lee JY, Jian CM, Xu C. 2023.
\textit{PRX Quantum} 4(3):030317

\bibitem{OgunnaikeFeldmeierLee2023}
{Ogunnaike} O, {Feldmeier} J, {Lee} JY. 2023.
\textit{Phys. Rev. Lett.} 131(22):220403

\bibitem{Moudgalya_2024}
Moudgalya S, Motrunich OI. 2024.
\textit{PRX Quantum} 5(4):040330

\bibitem{ChenGrover}
{Chen} YH, {Grover} T. 2024.
\textit{PRX Quantum} 5(3):030310

\bibitem{Maetal2025}
{Ma} R, {Zhang} JH, {Bi} Z, {Cheng} M, {Wang} C. 2025.
\textit{Physical Review X} 15(2):021062

\bibitem{Lessa2024}
{Lessa} LA, {Ma} R, {Zhang} JH, {Bi} Z, {Cheng} M, {Wang} C. 2025.
\textit{PRX Quantum} 6(1):010344

\bibitem{sala_spontaneous_2024}
Sala P, Gopalakrishnan S, Oshikawa M, You Y. 2024.
\textit{Physical Review B} 110(15):155150

\bibitem{BucaProsen2012}
{Bu{\v{c}}a} B, {Prosen} T. 2012.
\textit{New Journal of Physics} 14(7):073007

\bibitem{AlbertJiang2014}
Albert VV, Jiang L. 2014.
\textit{Phys. Rev. A} 89(2):022118

\bibitem{sang2024stability}
Sang S, Hsieh TH. 2025.
\textit{Phys. Rev. Lett.} 134(7):070403

\bibitem{hastings2011nonzero}
Hastings MB. 2011.
\textit{Physical review letters} 107(21):210501

\bibitem{coser2019classification}
Coser A, P{\'e}rez-Garc{\'\i}a D. 2019.
\textit{Quantum} 3:174

\bibitem{de2022symmetry}
de~Groot C, Turzillo A, Schuch N. 2022.
\textit{Quantum} 6:856

\bibitem{ma2023average}
Ma R, Wang C. 2023.
\textit{Phys. Rev. X} 13(3):031016

\bibitem{FanBaoMixedTopo}
Fan R, Bao Y, Altman E, Vishwanath A. 2024.
\textit{PRX Quantum} 5(2):020343

\bibitem{LeeYouXu2022}
{Lee} JY, {You} YZ, {Xu} C. 2022.
\textit{arXiv e-prints} :arXiv:2210.16323

\bibitem{ChenGrover2}
Chen YH, Grover T. 2024.
\textit{Phys. Rev. Lett.} 132(17):170602

\bibitem{sang2023mixed}
Sang S, Zou Y, Hsieh TH. 2024.
\textit{Phys. Rev. X} 14(3):031044

\bibitem{StableOpen2024}
Rakovszky T, Gopalakrishnan S, von Keyserlingk C. 2024.
\textit{Phys. Rev. X} 14(4):041031

\bibitem{lessa2025higher}
Lessa LA, Sang S, Lu TC, Hsieh TH, Wang C. 2025.
\textit{arXiv preprint arXiv:2503.12792}

\bibitem{Sang2025LR}
{Sang} S, {Lessa} LA, {Mong} RSK, {Grover} T, {Wang} C, {Hsieh} TH. 2025.
\textit{arXiv e-prints} :arXiv:2507.02292

\bibitem{Experiment2026}
{Wang} S, {Kiely} TG, {Tell} D, {Obermeyer} J, {Barendregt} M, et~al. 2026.
\textit{arXiv e-prints} :arXiv:2604.16137

\bibitem{Anderson1966}
{Anderson} PW. 1966.
\textit{Reviews of Modern Physics} 38(2):298--310

\bibitem{zhang2024fluctuation}
{Zhang} JH, {Xu} C, {Xu} Y. 2024.
\textit{arXiv e-prints} :arXiv:2409.18944

\bibitem{EdwardsAnderson}
Edwards SF, Anderson PW. 1975.
\textit{Journal of Physics F: Metal Physics} 5(5):965

\bibitem{Lessatoappear}
{Lessa} LA. 2026.
\textit{to appear}

\bibitem{Knap2025}
{Ziereis} N, {Moudgalya} S, {Knap} M. 2025.
\textit{arXiv e-prints} :arXiv:2509.09669

\bibitem{LiuRenyi1}
{Liu} Z, {Chen} L, {Zhang} Y, {Zhou} S, {Zhang} P. 2025.
\textit{Communications Physics} 8(1):274

\bibitem{WeinsteinRenyi1}
Weinstein Z. 2025.
\textit{Phys. Rev. Lett.} 134(15):150405

\bibitem{zhang2022strange}
{Zhang} JH, {Qi} Y, {Bi} Z. 2026.
{Fidelity strange correlators for average symmetry-protected topological
  phases}

\bibitem{Ma2024SPTdoubled}
Ma R, Turzillo A. 2025.
\textit{PRX Quantum} 6(1):010348

\bibitem{XueLeeBao}
{Xue} H, {Lee} JY, {Bao} Y. 2024.
\textit{arXiv e-prints} :arXiv:2403.17069

\bibitem{Guo2025}
Guo Y, Zhang JH, Zhang HR, Yang S, Bi Z. 2025.
\textit{Phys. Rev. X} 15(2):021060

\bibitem{Lu2024}
{Lu} S, {Zhu} P, {Lu} YM. 2024.
\textit{arXiv e-prints} :arXiv:2411.07174

\bibitem{Manjunath2025}
{Manjunath} N, {Turzillo} A, {Wang} C. 2025.
\textit{arXiv e-prints} :arXiv:2507.00127

\bibitem{BaoFandouble}
{Bao} Y, {Fan} R, {Vishwanath} A, {Altman} E. 2023.
\textit{arXiv e-prints} :arXiv:2301.05687

\bibitem{WangWuWang}
Wang Z, Wu Z, Wang Z. 2025.
\textit{PRX Quantum} 6(1):010314

\bibitem{RamanjitAbhinav}
Sohal R, Prem A. 2025.
\textit{PRX Quantum} 6(1):010313

\bibitem{EllisonCheng}
{Ellison} TD, {Cheng} M. 2025.
\textit{PRX Quantum} 6(1):010315

\bibitem{Li2026}
{Li} Z, {Firanko} R, {Hsieh} TH. 2026.
\textit{arXiv e-prints} :arXiv:2605.00088

\bibitem{Huangtoappear}
{Huang} X, {Sang} S, {Wang} C, {Zou} Y. 2026.
\textit{to appear}

\bibitem{LiuZou2026}
{Liu} Y, {Zou} Y. 2026.
\textit{arXiv e-prints} :arXiv:2601.13333

\bibitem{Kuno2024}
Kuno Y, Orito T, Ichinose I. 2024.
\textit{Phys. Rev. B} 110(9):094106

\bibitem{GuoYang2025}
Guo Y, Yang S. 2025.
\textit{Phys. Rev. B} 111(20):L201108

\bibitem{Shu2026}
{Shu} C, {Zhang} K, {Luo} ZX, {You} Y, {Sun} K. 2026.
\textit{arXiv e-prints} :arXiv:2603.06363

\bibitem{Ding2026}
{Ding} YM, {Guo} Y, {Bi} Z, {Yan} Z. 2026.
\textit{arXiv e-prints} :arXiv:2603.24342

\bibitem{Sarma2026}
{Sarma} A, {Xu} C. 2026.
\textit{arXiv e-prints} :arXiv:2603.24671

\bibitem{GilyenPoremba2022}
Gily{\'e}n A, Poremba A. 2024.
\textit{Quantum} 8:1222

\bibitem{Feng2025}
Feng X, Cheng Z, Ippoliti M. 2025.
\textit{Phys. Rev. Lett.} 135(20):200402

\bibitem{bratteli_operator_2012}
Bratteli O, Robinson DW. 2012.
Operator {{Algebras}} and {{Quantum Statistical Mechanics}}: {{Volume}} 1:
  {{C}}*- and {{W}}*- {{Algebras}}. {{Symmetry Groups}}. {{Decomposition}} of
  {{States}}.
Springer Science \& Business Media

\bibitem{nielsen_quantum_2010}
Nielsen MA, Chuang IL. 2010.
Quantum {Computation} and {Quantum} {Information}: 10th {Anniversary}
  {Edition}.
Cambridge: Cambridge University Press

\bibitem{Divi2026}
{Divi} F, {Lessa} LA, {Wang} C. 2026.
\textit{arXiv e-prints} :arXiv:2605.28967

\bibitem{LiuYiElse2026}
{Liu} R, {Yi} J, {Else} DV. 2026.
\textit{arXiv e-prints} :arXiv:2605.28925

\bibitem{Zhang2026}
{Zhang} C. 2026.
\textit{arXiv e-prints} :arXiv:2605.29113

\bibitem{ETHReview}
{D'Alessio} L, {Kafri} Y, {Polkovnikov} A, {Rigol} M. 2016.
\textit{Advances in Physics} 65(3):239--362

\bibitem{zhu2022nishimoris}
Zhu GY, Tantivasadakarn N, Vishwanath A, Trebst S, Verresen R. 2023.
\textit{Phys. Rev. Lett.} 131(20):200201

\bibitem{Lee2022}
{Lee} JY, {Ji} W, {Bi} Z, {Fisher} MPA. 2022.
\textit{arXiv e-prints} :arXiv:2208.11699

\bibitem{Nishimori1980}
Nishimori H. 1980.
\textit{Journal of Physics C: Solid State Physics} 13(21):4071--4076

\bibitem{MerzChalker2003}
Merz F, Chalker JT. 2002{\natexlab{a}}.
\textit{Phys. Rev. B} 66(5):054413

\bibitem{wang_analog_2024}
{Wang} TT, {Song} M, {Meng} ZY, {Grover} T. 2025{\natexlab{a}}.
\textit{PRX Quantum} 6(1):010358

\bibitem{merz2002rbim}
Merz F, Chalker J. 2002{\natexlab{b}}.
\textit{Physical Review B} 65(5):054425

\bibitem{RBIM1}
Honecker A, Picco M, Pujol P. 2001.
\textit{Phys. Rev. Lett.} 87(4):047201

\bibitem{RBIM2}
Picco M, Honecker A, Pujol P. 2006.
\textit{Journal of Statistical Mechanics: Theory and Experiment}
  2006(09):P09006

\bibitem{RBIM3}
Hasenbusch M, Toldin FP, Pelissetto A, Vicari E. 2008.
\textit{Phys. Rev. E} 77(5):051115

\bibitem{TQM}
{Dennis} E, {Kitaev} A, {Landahl} A, {Preskill} J. 2002.
\textit{Journal of Mathematical Physics} 43(9):4452--4505

\bibitem{WangPreskill2003}
{Wang} C, {Harrington} J, {Preskill} J. 2003.
\textit{Annals of Physics} 303(1):31--58

\bibitem{KogutReview}
Kogut JB. 1979.
\textit{Rev. Mod. Phys.} 51(4):659--713

\bibitem{GuoYangYu}
{Guo} Y, {Yang} S, {Yu} XJ. 2025.
\textit{PRX Quantum} 6(4):040311

\bibitem{Lin2025}
{Shah} J, {Fechisin} C, {Wang} YX, {Iosue} JT, {Watson} JD, et~al. 2025.
\textit{Quantum} 9:1912

\bibitem{SYK1}
{Kim} J, {Altman} E, {Lee} JY. 2024.
\textit{arXiv e-prints} :arXiv:2410.24225

\bibitem{SYK2}
{Chen} L, {Sun} N, {Zhang} P. 2025.
\textit{Phys. Rev. B} 111(6):L060304

\bibitem{MICorr2008}
Wolf MM, Verstraete F, Hastings MB, Cirac JI. 2008.
\textit{Phys. Rev. Lett.} 100(7):070502

\bibitem{LevinWen}
{Levin} M, {Wen} XG. 2006.
\textit{Phys. Rev. Lett.} 96(11):110405

\bibitem{Recoverability}
{Fawzi} O, {Renner} R. 2015.
\textit{Communications in Mathematical Physics} 340(2):575--611

\bibitem{Petz1986}
Petz D. 1986.
\textit{Communications in Mathematical Physics} 105(1):123--131

\bibitem{Chen2025}
{Chen} CF, {Rouz{\'e}} C. 2025.
\textit{arXiv e-prints} :arXiv:2504.02208

\bibitem{Kato2016}
{Kato} K, {Brandao} FGSL. 2016.
\textit{arXiv e-prints} :arXiv:1609.06636

\bibitem{Kuwahara2025}
{Kuwahara} T. 2025.
\textit{Physical Review X} 15(4):041010

\bibitem{ChenGuWenLRE}
Chen X, Gu ZC, Wen XG. 2010.
\textit{Phys. Rev. B} 82(15):155138

\bibitem{Yi2026}
Yi J, Li K, Liu C, Li Z, Zou L. 2026.
\textit{Phys. Rev. Lett.} 136(11):116604

\bibitem{LessaChengWang}
Lessa LA, Cheng M, Wang C. 2025.
\textit{Phys. Rev. X} 15(1):011069

\bibitem{XuJian2025}
{Xu} Y, {Jian} CM. 2025.
\textit{Phys. Rev. B} 111(12):125128

\bibitem{YouOshikawa}
{You} Y, {Oshikawa} M. 2024.
\textit{Phys. Rev. B} 110(16):165160

\bibitem{gaiotto_generalized_2014}
Gaiotto D, Kapustin A, Seiberg N, Willett B. 2015.
\textit{Journal of High Energy Physics} 2015(2):1--62

\bibitem{GeneralizedStatistics}
{Kobayashi} R, {Li} Y, {Xue} H, {Hsin} PS, {Chen} YA. 2026.
\textit{Physical Review X} 16(1):011010

\bibitem{Zhang2025}
Zhang C, Xu Y, Zhang JH, Xu C, Bi Z, Luo ZX. 2025.
\textit{Phys. Rev. B} 111(11):115137

\bibitem{MixedBootstrap2025}
{Yang} TH, {Shi} B, {Lee} JY. 2025.
\textit{arXiv e-prints} :arXiv:2506.04221

\bibitem{Lloyd2025}
{Lloyd} J, {Abanin} DA, {Gopalakrishnan} S. 2025.
\textit{arXiv e-prints} :arXiv:2508.02567

\bibitem{ChenGrover2025}
{Chen} YH, {Grover} T. 2025.
\textit{arXiv e-prints} :arXiv:2512.07220

\bibitem{LuHolography2025}
{Lu} TC, {Liu} YJ, {Gopalakrishnan} S, {You} Y. 2025.
\textit{arXiv e-prints} :arXiv:2511.19597

\bibitem{U1SWSSB1}
{Temkin} V, {Weinstein} Z, {Fan} R, {Podolsky} D, {Altman} E. 2025.
\textit{arXiv e-prints} :arXiv:2512.22119

\bibitem{U1Hydro}
{Hauser} J, {Su} K, {Ha} H, {Lloyd} J, {Kiely} TG, et~al. 2026.
\textit{arXiv e-prints} :arXiv:2602.16045

\bibitem{Vijay2025BKT}
{Vijay} A, {Lee} JY. 2025{\natexlab{a}}.
\textit{arXiv e-prints} :arXiv:2512.22121

\bibitem{Watanabe2024}
Watanabe H, Katsura H, Lee JY. 2024.
\textit{Phys. Rev. Lett.} 133(17):176001

\bibitem{GuWangWang2025}
{Gu} D, {Wang} Z, {Wang} Z. 2025.
\textit{Phys. Rev. B} 112(24):245123

\bibitem{HydroHuang}
Huang X, Qi M, Zhang JH, Lucas A. 2025.
\textit{Phys. Rev. B} 111(12):125147

\bibitem{Lee2026}
{Lee} JY. 2026.
\textit{arXiv e-prints} :arXiv:2605.05288

\bibitem{Livine2005}
Livine ER, Terno DR. 2005.
\textit{Phys. Rev. A} 72(2):022307

\bibitem{Sahu2024SU2}
Moharramipour A, Lessa LA, Wang C, Hsieh TH, Sahu S. 2024.
\textit{PRX Quantum} 5(4):040336

\bibitem{Li2025SU2}
Li Y, Pollmann F, Read N, Sala P. 2025.
\textit{Phys. Rev. X} 15(1):011068

\bibitem{WangLi2025}
{Wang} Z, {Li} L. 2025.
\textit{PRX Quantum} 6(1):010347

\bibitem{Sahu2026}
{Sahu} S, {Li} Y, {Sala} P. 2026.
\textit{Phys. Rev. A} 113(2):022406

\bibitem{Negari2025}
{Negari} AR, {Lessa} LA, {Sahu} S. 2025.
\textit{arXiv e-prints} :arXiv:2508.20166

\bibitem{Translation1}
{Lessa} LA, {Lu} TC. 2026.
\textit{arXiv e-prints} :arXiv:2605.15201

\bibitem{Translation2}
{Thorngren} R, {Gioia} L, {Zhang} C. 2026.
\textit{arXiv e-prints} :arXiv:2605.15200

\bibitem{Wang2021mixed}
{Shokrian Zini} M, {Wang} Z. 2021.
\textit{arXiv e-prints} :arXiv:2110.13946

\bibitem{LiLeeYoshida2025}
{Li} Z, {Lee} D, {Yoshida} B. 2025.
\textit{Physical Review X} 15(2):021090

\bibitem{Grover2025}
{Wang} TT, {Song} M, {Meng} ZY, {Grover} T. 2025{\natexlab{b}}.
\textit{PRX Quantum} 6(1):010358

\bibitem{HsinKobayashiRem2025}
{Hsin} PS, {Kobayashi} R, {Prem} A. 2025.
\textit{arXiv e-prints} :arXiv:2504.10569

\bibitem{SongZhang2025}
{Song} Z, {Zhang} JH. 2025.
\textit{arXiv e-prints} :arXiv:2509.24179

\bibitem{Sun2025}
Sun S, Zhang JH, Bi Z, You Y. 2025.
\textit{PRX Quantum} 6(2):020333

\bibitem{SymTFT1}
{Schafer-Nameki} S, {Tiwari} A, {Warman} A, {Zhang} C. 2025.
\textit{arXiv e-prints} :arXiv:2507.05350

\bibitem{SymTFT2}
{Qi} M, {Sohal} R, {Chen} X, {Stephen} DT, {Prem} A. 2025.
\textit{arXiv e-prints} :arXiv:2507.05335

\bibitem{SymTFT3}
{Luo} R, {Wang} YN, {Bi} Z. 2025.
\textit{PRX Quantum} 6(4):040358

\bibitem{Vijay2025Holo}
{Vijay} A, {Lee} JY. 2025{\natexlab{b}}.
\textit{arXiv e-prints} :arXiv:2511.21685

\bibitem{Sharpening}
{Agrawal} U, {Zabalo} A, {Chen} K, {Wilson} JH, {Potter} AC, et~al. 2022.
\textit{Physical Review X} 12(4):041002

\bibitem{Learnability2026}
Singh H, Vasseur R, Potter AC, Gopalakrishnan S. 2026.
\textit{Phys. Rev. B} 113(5):054305

\bibitem{Zerba2025}
{Zerba} C, {Gopalakrishnan} S, {Knap} M. 2025.
\textit{arXiv e-prints} :arXiv:2512.14830

\bibitem{Unlearnable}
{Advaith Kumar} T, {Zou} Y, {Negari} AR, {Melko} RG, {Hsieh} TH. 2026.
\textit{arXiv e-prints} :arXiv:2602.11262

\end{thebibliography}

\end{document}